\documentclass[aps,prb,superscriptaddress,twocolumn,floatfix]{revtex4-1}
\usepackage[abs]{overpic}
\usepackage{graphicx,graphics}
\usepackage{dcolumn}
\usepackage{latexsym,verbatim}
\usepackage{bm}
\usepackage{color}
\usepackage{ulem}
\usepackage[framemethod=default]{mdframed}
\usepackage[breaklinks=true,colorlinks,citecolor=blue,linkcolor=blue,urlcolor=blue]{hyperref}
\usepackage[makeroom]{cancel}

\usepackage{fancybox}
\begin{document}

\author{Alessandro Principi}
\affiliation{School of Physics and Astronomy, University of Manchester, Oxford Road, M13 9PL, Manchester,~UK}
\author{Denis Bandurin}
\affiliation{School of Physics and Astronomy, University of Manchester, Oxford Road, M13 9PL, Manchester,~UK}
\affiliation{Moscow Institute of Physics and Technology (State University), Dolgoprudny 141700,~Russia}
\author{Habib Rostami}
\affiliation{Nordita, KTH Royal Institute of Technology and Stockholm University, Stockholm, SE-106 91,~Sweden}
\author{Marco Polini}
\affiliation{Istituto Italiano di Tecnologia, Graphene Labs, Via Morego 30, I-16163 Genova,~Italy}
\affiliation{School of Physics and Astronomy, University of Manchester, Oxford Road, M13 9PL, Manchester,~UK}

\title{Pseudo-Euler equations from nonlinear optics: plasmon-assisted photodetection beyond hydrodynamics}
\begin{abstract}
A great deal of theoretical and experimental efforts have been devoted in the last decades to the study of long-wavelength photodetection mechanisms in field-effect transistors hosting two-dimensional (2D) electron systems. A particularly interesting subclass of these mechanisms is intrinsic and based on the conversion of the incoming electromagnetic radiation into plasmons, which resonantly enhance the photoresponse, and subsequent rectification via hydrodynamic nonlinearities. In this Article we show that such conversion and subsequent rectification occur well beyond the frequency regime in which hydrodynamic theory applies. We consider the nonlinear optical response of generic 2D electron systems and derive pseudo-Euler equations of motion for suitable collective variables. These are solved in one- and two-dimensional geometries for the case of graphene and the results are compared with those of hydrodynamic theory. Significant qualitative differences are found, which are amenable to experimental studies. Our theory expands the knowledge of the fundamental physics behind long-wavelength photodetection. 
\end{abstract}
\maketitle

\section{Introduction}
\label{sect:intro}

In a series of seminal papers~\cite{dyakonov_prl_1993, dyakonov_prb_1995, dyakonov_ieee_1996a, dyakonov_ieee_1996b}, which appeared in the mid nineties, Dyakonov and Shur (DS) proposed a very elegant mechanism that yields a finite dc response to an oscillating radiation field. Technologically, this is clearly extremely helpful since it means that one can detect rapidly oscillating electromagnetic fields (e.g. Terahertz fields) by carrying out a dc measurement.

The DS photodetection mechanism is based on the fact that a field-effect transistor (FET) hosting a two-dimensional (2D) electron gas (EG) acts as a cavity for plasma waves. (Plasma waves are collective oscillations that occur in a gated 2DEG, whereby the long-range tail of the Coulomb interaction among electrons is screened by the presence of a metal gate.)
When these are weakly damped, i.e.~when a plasma wave launched at the source can reach the drain in a time shorter than the momentum relaxation time $\tau$, the detection of radiation exploits constructive interference of the plasma waves in the cavity, which results in a resonantly enhanced response.
This is the so-called resonant regime of plasma-wave photodetection.

DS showed~\cite{dyakonov_ieee_1996a} that the photovoltage response of the 2DEG in a FET, i.e.~the electric potential difference between drain and source, contains a dc component even if the incoming field is ac, and thus provides rectification of the signal.
In the resonant regime, the dc photoresponse is characterized by peaks at odd multiples of the fundamental plasma-wave frequency.
This rectification mechanism is intrinsic, i.e.~it is not related to other rectification mechanisms (occurring, for the example, at the contacts) which could also be present in a real device.
Note that rectification of the signal is necessary to detect incoming radiation that exceeds the typical cutoff frequencies of circuit elements.
The DS mechanism is therefore particularly useful to detect Terahertz (THz) radiation.

The DS mechanism relies on the following two facts:
\begin{itemize}
\item[1)] The reflection symmetry corresponding to the exchange of source with drain in the FET channel is broken by the DS boundary conditions.
These boundary conditions are unusual because DS fixed the value of the current at the drain and the value of the potential at the source (instead of operating the device by fixing the current or the potential both at the source and at the drain, as is more customary).

\item[2)] The fact that the photovoltage averaged over a cycle of the oscillating radiation field is finite ultimately stems from the nonlinearity of the equations of motion describing transport in the FET channel. DS used the continuity and Navier-Stokes equations. When only the nonlinear terms in the former are considered (while the equation of motion for the velocity is purely of the Drude linear form), rectification arises as a result of the gate modulating {\it both} the electron density and the drift velocity in the channel. The Navier-Stokes equation~\cite{Landau06,falkovich_book}, however, strictly speaking is valid in a crystal when two conditions are satisfied. First, the mean free path for electron-electron (e-e) collisions $\ell_{\rm ee}$ needs to be much smaller than the device size $L$ and the mean free path for momentum-non-conserving collisions, $\ell$, i.e.~the following inequality must hold true: $\ell_{\rm ee}\ll \ell,L$. Second, during a cycle of oscillation of the electromagnetic field, electrons need to have enough time to reach a state of local thermal equilibrium, i.e.~the following inequality must also hold true: $\omega \tau_{\rm ee} \ll 1$, where $\omega$ is the angular frequency of the external field and $\tau_{\rm ee} = \ell_{\rm ee}/v_{\rm F}$ is the mean free time for e-e collisions, $v_{\rm F}$ being the Fermi velocity.
\end{itemize}
Despite the solid experimental evidence of hydrodynamic flow in solid-state devices that has recently appeared in the literature in a variety of material systems~\cite{bandurin_science_2016,kumar_natphys_2017,berdyugin_arxiv_2018,bandurin_arxiv_2018,moll_science_2016,crossno_science_2016,braem_arxiv_2018}, DS theory is limited in that it puts a severe constraint on the range of frequencies of the external electromagnetic field---see point 2) above. For example, for the case of graphene, $\tau_{\rm ee}$ at typical carrier densities ($n=1.0 \times 10^{12}~{\rm cm}^{-2}$, say) and  temperature $T=300~{\rm K}$ is on the order of~\cite{principi_prb_2016,otherparameters} $0.15~{\rm ps}$ and therefore hydrodynamic theory is strictly applicable at finite $\omega$ only for frequencies $\nu = \omega/(2\pi)\ll 1~{\rm THz}$.

In this Article we transcend DS theory in that we do not assume that the above hydrodynamic inequalities hold true. In full generality, we simply acknowledge that the 2DEG in the FET channel has a frequency-dependent nonlinear optical response of intrinsic but unspecified origin, parametrized via nonlinear conductivity tensors. From the nonlinear optical response of the 2DEG we derive a generalized Euler equation (which we term ``pseudo-Euler'' equation) of motion for the velocity field ${\bm v}({\bm r},t)\equiv {\bm j}({\bm r},t)/n({\bm r},t)$, where $n({\bm r},t)$ and ${\bm j}({\bm r},t)$ are the number density and the charge current density flowing in the FET channel, respectively, which are related by the continuity equation. Physical parameters related to the particular material of which the FET channel is made, e.g.~GaAs, graphene, etc, enter our theory only through the nonlinear conductivity tensors.

As a concrete example of our theory, we focus on the specific case of graphene, restricting ourselves to second-order nonlinearities and using results available in the literature~\cite{Rostami_prb_2017} for the second-order conductivity tensor $\sigma^{(2)}_{ijk}({\bm q}, {\bm q}_1, {\bm q}_2, \omega, \omega_1, \omega_2)$. In this case, we derive and solve the pseudo-Euler equation of motion that emerges for ${\bm v}({\bm r},t)$ and discuss qualitative and quantitative differences with respect to the canonical Euler equation of hydrodynamic theory.

Our Article is organized as follows. In Sect.~\ref{sect:model} we present a derivation of a general pseudo-Euler equation of motion from nonlinear optics. In Sect.~\ref{sect:graphene} we apply our theory to graphene. Analytical and numerical results for the related dc photoresponse in the case of one- and two-dimensional geometries are presented in Sects.~\ref{sect:photo_1D} and~\ref{sect:phot_2D}, respectively. A  summary and a brief set of conclusions are reported in Sect.~\ref{sect:summary}. Relevant details are included in two Appendices. In Appendix~\ref{sect:hydro_cond_euler} we show that the canonical Euler equation of hydrodynamic theory can be derived from our general theory when the hydrodynamic second-order nonlinear conductivity tensor~\cite{sun_pnas_2018} is used. Finally, in Appendix~\ref{app:simplification_nonlinear} we summarize the main algebraic steps that are needed to simplify the last term of Eq.~(\ref{eq:cont_NS_nonlinear_2}).

\section{Derivation of the pseudo-Euler equation from nonlinear optics}
\label{sect:model}
To describe nonlinear electron flow in a FET channel hosting a generic 2DEG we employ a set of equations of motion for two collective variables, i.e.~the number density $n({\bm r}, t)$ and the charge current ${\bm j}({\bm r},t)$. These are related by the continuity equation
\begin{eqnarray} \label{eq:continuity}
- e \partial_t n({\bm r}, t) + {\bm \nabla}\cdot {\bm j}({\bm r},t) = 0 
~,
\end{eqnarray}
where $-e<0$ is the electron charge. When an electric field ${\bm E}_{\rm ext} ({\bm r}, t)$ is applied to the system, the current responds according to [see, e.g., Refs.~\onlinecite{Shen_book,Jackson_chap_6,cheng_scirep_2017}]
\begin{eqnarray} \label{eq:current_field}
j_{i}({\bm q},\omega) &=& \sum_{j} \sigma_{ij}^{(1)}({\bm q}, \omega) E_j ({\bm q},\omega) 
\nonumber\\
&+&
\frac{1}{2}\sum_{{\bm q}_1,{\bm q}_2} \sum_{\omega_1, \omega_2} \sum_{j,k} \sigma_{ijk}^{(2)}({\bm q}, {\bm q}_1, {\bm q}_2, \omega, \omega_1, \omega_2) 
\nonumber\\
&\times&
E_j ({\bm q}_1,\omega_1) E_k ({\bm q}_2,\omega_2) + \ldots
~.
\end{eqnarray}
Terms of higher-order in the electric field are not explicitly written down in Eq.~(\ref{eq:current_field}) but their inclusion is straightforward. 
In Eq.~(\ref{eq:current_field}), Latin indices $i,j,k=x,y$ denote the Cartesian components of the vectors ${\bm j}({\bm q},\omega)$ and ${\bm E} ({\bm q},\omega)$, which are the Fourier components of the current ${\bm j}({\bm r},t)$ and {\it total} electric field ${\bm E}({\bm r},t) = {\bm E}_{\rm ext}({\bm r},t) -{\bm \nabla} U({\bm r},t)$, respectively. Here $U({\bm r},t)$ is the ``gate-to-channel swing''~\cite{tomadin_prb_2013}
\begin{eqnarray} \label{eq:field_sc}
U({\bm r},t) = U_0 -e \int d{\bm r}' V({\bm r}-{\bm r}') \delta n({\bm r}',t)~.
\end{eqnarray}
In this equation $U_0$ is the potential at the metallic gate~\cite{tomadin_prb_2013}, $V({\bm r}-{\bm r}')$ is the appropriate e-e interaction, which depends on the details of the configuration of dielectrics and gates surrounding the 2DEG, whereas $\delta n({\bm r},t)$ is the deviation of the electronic density $n({\bm r},t)$ from its equilibrium value $n_0$. The latter is determined by the geometrical capacitance per unit area $C$ of the device (see also below) as $n_0 = -C U_0/e$. 
From the above discussion, it is clear that the quantities $\sigma_{ij}^{(1)}({\bm q}, \omega)$ and $\sigma_{ijk}^{(2)}({\bm q}, {\bm q}_1, {\bm q}_2, \omega, \omega_1, \omega_2)$ in Eq.~(\ref{eq:current_field}) are therefore the {\it proper}~\cite{Giuliani_and_Vignale} first- and second-order non-local conductivities, which can be calculated microscopically for a given 2DEG Hamiltonian (see, e.g., Refs.~\onlinecite{Rostami_prb_2017,cheng_scirep_2017,Mikhailov_prb_2011,jafari_jpcm_2012,mikhailov_prb_2014,Mikhailov_prl_2014,cheng_njp_2014,wehling_prb_2015,cheng_prb_2015,Habib_prb_2016,mikhailov_prb_2016,mikhailov_prb_2017,sun_naturecommun_2018,soavi_naturenano_2018}). These encode the response to the total electric field ${\bm E}({\bm r},t)$ which is the sum of ${\bm E}_{\rm ext}({\bm r},t)$ and $-\nabla U({\bm r},t)$.
Self-sustained collective excitations in the charge channel~\cite{Giuliani_and_Vignale} (e.g.~plasmons, plasma waves, etc) are described by the nontrivial solutions of the problem posed by Eqs.~(\ref{eq:continuity})-(\ref{eq:field_sc}) when ${\bm E}_{\rm ext}({\bm r},t)={\bm 0}$.
We stress that the description so far is completely general and applies to FETs made with any 2DEG and any configuration of gates and dielectrics. Information on the material forming the FET channel enters the problem only through the expressions of the proper conductivities $\sigma_{ij}^{(1)}({\bm q}, \omega)$, $\sigma_{ijk}^{(2)}({\bm q}, {\bm q}_1, {\bm q}_2, \omega, \omega_1, \omega_2)$, etc.

We now come to the crucial point of our theory. To make contact with the standard DS formulation, we recast Eq.~(\ref{eq:current_field}) in the form of generalized Euler equation of motion. This task is achieved by inverting Eq.~(\ref{eq:current_field}) order-by-order in the nonlinearities, rewriting it as 
\begin{eqnarray} \label{eq:general_E_j_rel} 
E_{i}(\Omega) &=& \rho_{ij}^{(1)}(\Omega) j_j (\Omega) - \frac{1}{2} \rho_{ijk}^{(2)}(\Omega, \Omega_1, \Omega_2) j_j (\Omega_1) j_k (\Omega_2) 
\nonumber\\
&+& \ldots
~.
\end{eqnarray}
To lighten the notation, we introduced frequency-momentum variables $\Omega_i \equiv(\omega_i, {\bm q}_i)$ and adopted Einstein's summation convention on repeated indices, as well as on $\Omega_1$ and $\Omega_2$. The minus sign in front of the second term on the right-hand side of Eq.~(\ref{eq:general_E_j_rel}) has been introduced for convenience. Eq.~(\ref{eq:general_E_j_rel}) is the most important result of this Section. A generalized Euler differential equation can be obtained by finding the coefficients $\rho_{ij}^{(1)}$, $\rho_{ijk}^{(2)}$, etc., and Fourier transforming to real space and time.

The quantities $\rho_{ij}^{(1)}$, $\rho_{ijk}^{(2)}$, etc. can be found as following. We first rewrite Eq.~(\ref{eq:current_field}) as
\begin{eqnarray} \label{eq:general_j_E_rel}
j_{i}(\Omega) &=& \lambda \sigma_{ij}^{(1)}(\Omega) E_j (\Omega) +\nonumber\\
&+& \frac{\lambda^2}{2} \sigma_{ijk}^{(2)}(\Omega, \Omega_1, \Omega_2) E_j (\Omega_1) E_k (\Omega_2) 
\nonumber\\
&+&
\ldots
~.
\end{eqnarray}
Here, we momentarily introduced the book-keeping dimensionless parameter $\lambda$, which we use to keep track of the order of nonlinearity and we will set to one once inversion is performed. By plugging Eq.~(\ref{eq:general_E_j_rel}) into Eq.~(\ref{eq:general_j_E_rel}), and collecting terms containing the first, second, etc. power of $\lambda$ we find
\begin{eqnarray} \label{eq:sigma_collect_1}
\sigma_{im}^{(1)}(\Omega) \rho_{mj}^{(1)}(\Omega) = \delta_{ij}
\end{eqnarray}
and
\begin{eqnarray} \label{eq:sigma_collect_2}
&&\sigma_{ij}^{(1)}(\Omega) \rho_{jmn}^{(2)}(\Omega, \Omega_1, \Omega_2) = \nonumber\\
&&\sigma_{i\ell k}^{(2)}(\Omega, \Omega_1, \Omega_2) \rho_{\ell m}^{(1)}(\Omega_1) \rho_{kn}^{(1)}(\Omega_2)~,
\end{eqnarray}
and so on. Eqs.~(\ref{eq:sigma_collect_1})-(\ref{eq:sigma_collect_2}) can be solved iteratively, and give
\begin{eqnarray} \label{eq:linear_resistivity_general}
\rho_{ij}^{(1)}(\Omega) = \big[\sigma^{(1)}(\Omega)\big]^{-1}_{ij}
\end{eqnarray}
and
\begin{eqnarray} \label{eq:nonlinear_resistivity_general}
&&\rho_{ijk}^{(2)}(\Omega, \Omega_1, \Omega_2) = \nonumber\\
&&\sigma_{mn\ell}^{(2)}(\Omega, \Omega_1, \Omega_2) \rho_{mi}^{(1)}(\Omega) \rho_{nj}^{(1)}(\Omega_1) \rho_{\ell k}^{(1)}(\Omega_2)~,
\end{eqnarray}
and so on. 

As a reality check, in Appendix~\ref{sect:hydro_cond_euler} we show that the procedure outlined in this Section leads to the canonical Euler equation when the hydrodynamic nonlinear conductivity tensor is used for $\sigma^{(2)}_{ijk}$.

\section{Pseudo-Euler equation from the ballistic second-order conductivity: the case of graphene}
\label{sect:graphene}
In general~\cite{Shen_book,Jackson_chap_6}, slowly-varying currents induced in a medium by an electric field can be related to the derivatives of the electric polarizability ${\bm P}(t)$, the electric quadrupole tensor ${\bm Q}({\bm r},t)$, and the magnetic dipole moment ${\bm M}({\bm r},t)$. Defining the {\it local} polarization current, ${\bm j}_{\rm P}(t) = \partial_t {\bm P}(t)$, and the {\it non-local} quadrupole and magnetization currents, $ {\bm j}_Q({\bm r},t) \propto \partial_t {\bm \nabla} \cdot {\bm Q}({\bm r},t)$ and ${\bm j}_{\rm M}({\bm r},t) \propto {\bm \nabla}\times {\bm M}({\bm r},t)$, we can write the total current as ${\bm j}({\bm r},t) \approx {\bm j}_{\rm P}(t)+{\bm j}_{\rm Q}({\bm r},t)+{\bm j}_{\rm M}({\bm r},t)$. This expression is valid for currents which vary slowly in both space and time, for which higher-order derivatives can be neglected. Each of the three currents, ${\bm j}_{\rm P}(t)$, ${\bm j}_{\rm Q}({\bm r},t)$ and ${\bm j}_{\rm M}({\bm r},t)$ is a nonlinear function of the external electric field.

To linear order in the applied field, and in the absence of a magnetic field, ${\bm j}^{(1)}_{\rm P}(t)$ and ${\bm j}^{(1)}_{\rm Q}({\bm r},t)$ are finite, whereas ${\bm j}^{(1)}_{\rm M}({\bm r},t)$ vanishes~\cite{Giuliani_and_Vignale}. Since ${\bm j}^{(1)}_{\rm P}(t)$ is finite, when considering long-wavelength properties of the system we can neglect ${\bm j}^{(1)}_{\rm Q}({\bm r},t)$. 

In centro-symmetric systems such as pristine graphene, the nonlinear (second-order) polarization current ${\bm j}^{(2)}_{\rm P}(t)= \partial_t {\bm P}^{(2)}(t)$ vanishes by symmetry. Therefore, the first nontrivial contribution to ${\bm j}^{(2)}({\bm r},t)$ is due to the {\it non-local} ${\bm j}^{(2)}_{\rm Q}({\bm r},t)$ and ${\bm j}^{(2)}_{\rm M}({\bm r},t)$ originating from the second-order electric quadrupole tensor ${\bm Q}^{(2)}({\bm r},t)$ and magnetic dipole moment $\bm M^{(2)}({\bm r},t)$. Note that ${\bm j}^{(2)}_{\rm M}({\bm r},t)$ is in general nonzero: magnetic fields are generated through the curl of the non-uniform electric field [see Eq.~(\ref{eq:j_E_rel_non_hydro}) below]. Therefore, the total current is ${\bm j}({\bm r},t) \approx {\bm j}^{(1)}_{\rm P}(t)+{\bm j}^{(2)}_{\rm Q}({\bm r},t)+{\bm j}^{(2)}_{\rm M}({\bm r},t)$.
From this expression, it is then clear that the leading contribution to the second-order conductivity of a centro-symmetric system is linear in the wave vector~\cite{Rostami_prb_2017,cheng_scirep_2017,Wang_prb_2016}, {\it i.e.}
\begin{eqnarray} \label{eq:nonlinear_cond_ds}
\sigma^{(2)}_{ijk}(\Omega, \Omega_1, \Omega_2) &=& \delta_{\Omega_1+\Omega_2,\Omega} \sum_{\beta}\Big\{ q_{1,\beta} d_{ijk\beta}(\omega,\omega_1) 
\nonumber\\
&+&
 q_{2,\beta} d_{ikj\beta}(\omega,\omega_2) \Big \}~,
\end{eqnarray} 
where $d_{ijk\beta}(\omega,\omega_1)$ is a rank-4 tensor. Following the discussion above, $d_{ijk\beta}(\omega,\omega_1)$ can be split into electric-quadrupole and magnetic-dipole contributions, denoted by $d^{(\rm Q)}_{ijk\beta}(\omega,\omega_1)$ and $d^{(\rm M)}_{ijk\beta}(\omega,\omega_1)$, respectively. 
Note that Eq.~(\ref{eq:nonlinear_cond_ds}) makes explicit the permutation symmetry $\sigma^{(2)}_{ijk}(\Omega, \Omega_1, \Omega_2)=\sigma^{(2)}_{ikj}(\Omega, \Omega_2, \Omega_1)$.  

In the case of graphene (D$_{6\rm h}$ point group symmetry), only a few tensor elements are independent, i.e.~$d_{xxyy}(\omega,\omega_1)$, $d_{xyxy}(\omega,\omega_1)$, $d_{xyyx}(\omega,\omega_1)$ and $d_{xxxx}(\omega,\omega_1)=d_{xxyy}(\omega,\omega_1)+d_{xyxy}(\omega,\omega_1)+d_{xyyx}(\omega,\omega_1)$. 
Further permutation symmetries discussed in Ref.~[\onlinecite{Rostami_prb_2017}] and Ref.~[\onlinecite{cheng_scirep_2017}]  lead to the symmetry relation $d^{(\rm Q)}_{xyyx}(\omega,\omega_1)=d^{(\rm Q)}_{xxyy}(\omega,\omega_1)$ and to the vanishing of all $d^{(\rm M)}_{ijk\beta}(\omega,\omega_1)$ except for $d^{(\rm M)}_{xxyy}(\omega,\omega_1)$. Considering all these symmetry constrains we get~\cite{cheng_scirep_2017} 
\begin{eqnarray} \label{eq:j_E_rel_non_hydro}
&&{\bm j}(\Omega) = \sigma^{(1)}(\omega) {\bm E} (\Omega) \nonumber \\
&+&\sum_{\Omega_1} \Big\{ 2 d^{({\rm Q})}_{xxyy}(\omega,\omega_1) \big[ {\bm E} (\Omega_2)\cdot {\bm q}_1 \big] {\bm E} (\Omega_1)
\nonumber\\&+& \big[d^{({\rm M})}_{xxyy}(\omega,\omega_1) - d^{({\rm Q})}_{xxyy}(\omega,\omega_1)\big] \big[{\bm q}_1\times {\bm E} (\Omega_1)\big] \times {\bm E} (\Omega_2)
\nonumber\\
&+& 
d^{({\rm Q})}_{xyxy}(\omega,\omega_1) \big[ {\bm q}_1\cdot {\bm E} (\Omega_1) \big] {\bm E} (\Omega_2)
\Big\}~.
\end{eqnarray}
In Eq.~(\ref{eq:j_E_rel_non_hydro}), $\sigma^{(1)}(\omega)$ is the first-order local conductivity which can be calculated, as customary, from the continuum massless-Dirac-fermion (MDF) model~\cite{katsnelson_book}. Hereafter, we drop the superscript ``$(1)$'' in $\sigma^{(1)}(\omega)$ to lighten the notation, and we set $\hbar =1$. In the MDF model, $\sigma(\omega) = i n_0 e^2 f(\omega)/(m \omega)$, where $n_0$ is the equilibrium density, $m = k_{\rm F}/v_{\rm F}$ the effective mass, $k_{\rm F} = \sqrt{4\pi n_0/N_{\rm F}}$ the Fermi wave number, $N_{\rm F}=4$  the number of fermion flavors, $v_{\rm F}\approx 1.0 \times 10^6~{\rm m}/{\rm s}$ the Fermi velocity, and, finally, the function $f(\omega)$ accounts for the contribution of inter-band transitions for $\omega<2E_{\rm F}$:
\begin{equation} \label{eq:linear_conductivity_def}
f(\omega) = 1 + \frac{m \omega}{16 \pi n_0} \ln\left| \frac{2 E_{\rm F} - \omega}{2 E_{\rm F} + \omega} \right|
~.
\end{equation}
Such logarithmic singularity in the imaginary part of the linear conductivity is due to Pauli blocking. Since interband transitions are forbidden (allowed) for $\omega < 2 E_{\rm F}$ ($\omega > 2 E_{\rm F}$), the {\it real} part of the linear conductivity must jump from zero to a finite value at $\omega = 2 E_{\rm F}$. Peculiar to graphene, the real part of the linear conductivity is also constant for $\omega > 2 E_{\rm F}$, {\it i.e.} $\Re e\sigma^{(1)}(\omega) \propto \Theta(\omega - 2E_{\rm F})$.~\cite{katsnelson_book} Its {\it imaginary} part is obtained by a Kramers-Kronig transform~\cite{Giuliani_and_Vignale} and, given the form of $\Re e\sigma^{(1)}(\omega)$, must exhibit a logarithmic singularity at $\omega = 2E_{\rm F}$. Note that the singularity is peculiar to the zero-temperature limit, and is cured by thermal fluctuations.

Similarly, the calculation of the electric quadrupole and magnetic dipole contributions to $d_{ijk\beta}(\omega,\omega_1)$~\cite{Rostami_prb_2017,cheng_scirep_2017,Wang_prb_2016} within the MDF model yields~\cite{katsnelson_book}:
\begin{eqnarray}\label{eq:d2xyyx_main}
d^{({\rm Q})}_{xxyy}(\omega, \omega_1)=
\frac{8 E^2_{\rm F} d_0}{\omega_1^2\omega_2 \omega}\frac{ \omega_1^2 \omega -2\omega_2 E^2_{\rm F} }{\left [\omega_1^2-4 E^2_{\rm F} \right ]  \left [ \omega^2 -4 E^2_{\rm F} \right ]}
~,
\end{eqnarray}
\begin{eqnarray}\label{eq:d2xyxy_main}
d^{({\rm Q})}_{xyxy}(\omega, \omega_1) =  
\frac{16 E^2_{\rm F} d_0}{\omega_1^2\omega_2 \omega}\frac{ E^2_{\rm F} (2\omega_1+3\omega_2)-\omega_1^2 \omega }{\left [\omega_1^2-4 E^2_{\rm F} \right ]  \left [ \omega^2 -4 E^2_{\rm F} \right ]}
~,
\end{eqnarray}
and
\begin{eqnarray}\label{eq:d2xyyxM_main}
d^{({\rm M})}_{xxyy}(\omega, \omega_1)=  
\frac{8 E_{\rm F}^2 d_0}{\omega_1 \omega_2 \omega} 
\frac{ \omega  (2 \omega + \omega_1)-4 E_{\rm F}^2 }{\big[\omega^2-4 E_{\rm F}^2\big] \big[ \omega_1^2 -4 E_{\rm F}^2 \big]}~.
\end{eqnarray}
Here, $d_0 = n_0 e^3/m^2$. 
In passing, we note that it is possible to calculate the electric-quadrupole $d^{({\rm Q})}_{ijk\beta}(\omega, \omega_1)$ contributions in the scalar potential gauge~\cite{Rostami_prb_2017}, whereas the magnetic-dipole $d^{({\rm M})}_{ijk\beta}(\omega, \omega_1)$ contributions require the evaluation of the electrical response in the presence of an inhomogeneous vector potential~\cite{cheng_scirep_2017,Wang_prb_2016}.
The divergences in Eqs.~(\ref{eq:d2xyyx_main})-(\ref{eq:d2xyyxM_main}) have the same physical origin of the logarithmic singularity of Eq.~(\ref{eq:linear_conductivity_def}), {\it i.e.} Pauli blocking in the zero-temperature limit. Similarly, they are cured by the inclusion of thermal broadening. Note that the latter will not wash them out completely, since these singularities are too strong ({\it i.e.} $\propto 1/x$) to disappear in the presence of a sufficiently small broadening.
Thus, the enhancement of the DS rectified potential we discuss below, which relies on the singularities of Eqs.~(\ref{eq:d2xyyx_main})-(\ref{eq:d2xyyxM_main}), will be observable in a well defined temperature window.

Inverting Eq.~(\ref{eq:j_E_rel_non_hydro}), as explained in Sect.~\ref{sect:model}, we find
\begin{widetext}
\begin{eqnarray} \label{eq:euler_non_hydro_2}
-i \omega \sigma(\omega) {\bm E}(\Omega) &=& 
-i\omega {\bm j}(\Omega)
+
\sum_{\Omega_1} \frac{i\omega}{\sigma(\omega_1) \sigma(\omega_2)} \Big\{ \big[d^{({\rm M})}_{xxyy}(\omega,\omega_1) - d^{({\rm Q})}_{xxyy}(\omega,\omega_1)\big] \big[{\bm q}_1\times {\bm j} (\Omega_1)\big] \times {\bm j} (\Omega_2)
\nonumber\\
&+& 
2 d^{({\rm Q})}_{xxyy}(\omega,\omega_1) \big[ {\bm j} (\Omega_2)\cdot {\bm q}_1 \big] {\bm j} (\Omega_1)
+
d^{({\rm Q})}_{xyxy}(\omega,\omega_1) \big[ {\bm q}_1\cdot {\bm j} (\Omega_1) \big] {\bm j} (\Omega_2)
\Big\}~.
\end{eqnarray}
Fourier transforming Eq.~(\ref{eq:euler_non_hydro_2}) to real space we find the appropriate pseudo-Euler equation for a graphene FET channel:
\begin{eqnarray} \label{eq:euler_non_hydro_3}
-i \omega \sigma(\omega) {\bm E}({\bm r},\omega) &=& 
-i\omega {\bm j}({\bm r}, \omega)
+
\sum_{\omega_1} \frac{\omega}{\sigma(\omega_1) \sigma(\omega_2)} \Big\{ \big[d^{({\rm M})}_{xxyy}(\omega,\omega_1) - d^{({\rm Q})}_{xxyy}(\omega,\omega_1)\big] \big[{\bm \nabla}\times {\bm j} ({\bm r},\omega_1)\big] \times {\bm j} ({\bm r},\omega_2)
\nonumber\\
&+& 
2 d^{({\rm Q})}_{xxyy}(\omega,\omega_1) \big[ {\bm j} ({\bm r},\omega_2)\cdot {\bm \nabla} \big] {\bm j} ({\bm r},\omega_1)
+
d^{({\rm Q})}_{xyxy}(\omega,\omega_1) \big[ {\bm \nabla}\cdot {\bm j} ({\bm r},\omega_1) \big] {\bm j} ({\bm r},\omega_2)
\Big\}~.
\end{eqnarray}
\end{widetext}

We now consider the problem of rectification of radiation by the intrinsic nonlinearities of the system. We will study two cases:
\begin{itemize}
\item[i)] In the first one, following Refs.~\onlinecite{dyakonov_prl_1993, dyakonov_prb_1995, dyakonov_ieee_1996a, dyakonov_ieee_1996b}, we consider a one-dimensional (1D) geometry, i.e.~a gated FET channel where the source contact oscillates at frequency $\omega$ while the drain is left fluctuating (i.e.~no current flows through it). In this case, we generalize the DS theory to arbitrary frequencies, fully transcending the assumption $\omega\tau_{\rm ee}\ll 1$. Our theory reduces to the standard DS one in the low-frequency limit. In the high-frequency regime, instead, we find that the dc photoresponse is enhanced when the radiation frequency is close to the threshold of inter-band absorption, i.e.~when $\omega \to 2 E_{\rm F}$.

\item[ii)] In the second one, we study a 2D geometry, whereby oscillating or fluctuating contacts are connected to {\it all} edges of the channel and can be used to impose arbitrary boundary conditions. This geometry and its flexibility allow to highlight the impact of beyond-hydrodynamics terms contained in our pseudo-Euler equation~(\ref{eq:euler_non_hydro_3}). We find in fact that, even in the low-frequency limit ($\omega \ll 2E_{\rm F}$), beyond-hydrodynamic corrections to the canonical Euler equation produce new qualitative features in the dc photoresponse.
\end{itemize}
To calculate the rectified photoresponse of a FET channel, each quantity in Eqs.~(\ref{eq:continuity}) and~(\ref{eq:euler_non_hydro_3}) is expanded as ${\cal O}({\bm r},t) = {\cal O}_0 +{\cal O}_1({\bm r},t)+{\cal O}_2({\bm r},t)+\ldots$, where ${\cal O}_0$ is the equilibrium value, ${\cal O}_1({\bm r},t)$ is the deviation from the equilibrium value which is linear in the fields and currents, while ${\cal O}_2({\bm r},t)$ is the first nonlinear contribution. To simplify the calculations, we assume that a gate is placed in close proximity to the 2DEG, so that the density $n({\bm r},t)$ and the gate-to-channel swing $U({\bm r},t)$ are linearly related by the so-called local-capacitance approximation~\cite{dyakonov_prl_1993,tomadin_prb_2013}:
\begin{eqnarray} \label{eq:local_cap_approx}
U({\bm r},t) = - \frac{e}{C} n({\bm r},t)~,
\end{eqnarray}
i.e. we approximate~$V({\bm r}-{\bm r}') = \delta({\bm r}-{\bm r}')/C$ in Eq.~(\ref{eq:field_sc}). Here, $C$ is a suitable geometrical capacitance for the FET of interest, which will not be specified any further for the sake of generality.
The local capacitance approximation~(\ref{eq:local_cap_approx}) captures the relevant physics and agrees, at least qualitatively, with experiments. The presence of a metal gate, both in the theory and in experiments, screens the long-range part of the Coulomb interaction, making it effectively short-ranged. The interaction in the presence of the gate decays exponentially at large distances, with a typical length scale controlled by the distance between the channel and the gate ($d_{\rm g-c}$). The shorter the distance (in comparison to the typical wavelength of density oscillations $\lambda_{\rm p}$), the more short-ranged is the effective interaction. Eq.~(\ref{eq:local_cap_approx}) represents therefore the screened Coulomb interaction in the limit of negligible $d_{\rm g-c}/\lambda_{\rm p}$. A more complete discussion can be found in Ref.~\onlinecite{tomadin_prb_2013}.

The potential $U({\bm r},t)$ oscillates around its equilibrium value $U_0$, which is homogeneous, static and defines the equilibrium channel density $n_0 = - C U_0/e$, around which $n({\bm r},t)$ fluctuates.
We first solve the linear problem, to find the plasmon field $U_1({\bm r},t)$ and the electron velocity ${\bm v}_1({\bm r},t)$. These are then used to determine the nonlinear rectified potential, averaged over a period of oscillation, which will be denoted by $U_2({\bm r})$.

\section{Theory of photodetection in 1D geometries}
\label{sect:photo_1D}
We now consider a channel of finite length $L$ in the ${\hat {\bm x}}$ direction, with contacts located at $x=0$ and $x=L$, and infinitely long in the ${\hat {\bm y}}$ direction. In this case, all functions are independent of $y$ and the photodetection problem becomes effectively 1D. Using the local-capacitance relation~(\ref{eq:local_cap_approx}), Eqs.~(\ref{eq:continuity}) and~(\ref{eq:euler_non_hydro_3}) read
\begin{eqnarray} \label{eq:DS_continuity}
&&
C \partial_t U(x, t) + \partial_x j (x,t) = 0 
~,
\end{eqnarray}
and
\begin{eqnarray} \label{eq:DS_euler}
&&
-i \omega \sigma(\omega) E(x,\omega) = 
-i\omega j(x, \omega)
+
\sum_{\omega_1} \frac{\omega}{\sigma(\omega_1) \sigma(\omega_2)} 
\nonumber\\
&&
\times
d^{({\rm Q})}_{xxxx}(\omega,\omega_1) j (x,\omega_2)\partial_x  j (x,\omega_1)
~,
\end{eqnarray}
where~\cite{Rostami_prb_2017} $d^{({\rm Q})}_{xxxx}(\omega,\omega_1) = 2 d^{({\rm Q})}_{xxyy}(\omega, \omega_1) + d^{({\rm Q})}_{xyxy}(\omega, \omega_1)$ and $E(x,\omega) = -\partial_x U(x,\omega)$.
We now define
\begin{eqnarray}\label{eq:g_function}
g(\omega, \omega_1) = \frac{16 E_{\rm F}^4 }{\left [(\hbar\omega_1)^2-4 E^2_{\rm F} \right ]  \left [ (\hbar\omega)^2 -4 E^2_{\rm F} \right ]}~,
\end{eqnarray}
which allows to rewrite Eq.~(\ref{eq:DS_euler}) as 
\begin{eqnarray} \label{eq:DS_euler_2}
&&
\frac{n_0 e^2}{m} f(\omega) E(x,\omega) =
-i\omega j(x, \omega)
- \frac{1}{n_0 e}
\sum_{\omega_1} \frac{g(\omega,\omega_1) }{f(\omega_1) f(\omega_2)} 
\nonumber\\
&&
\times
\big[
j(x,\omega_2)\partial_x  j (x,\omega_1)
- i \omega e j(x,\omega_2)\delta n (x,\omega_1)
\big]
~.
\end{eqnarray}
Recall that the function $f(\omega)$ has been defined in Eq.~(\ref{eq:linear_conductivity_def}), and that $\delta n (x,\omega)$ is related to $U (x,\omega) - U_0$ via Eq.~(\ref{eq:local_cap_approx}).  
To obtain an explicit Euler-like equation, we now introduce the velocity field $v(x,t)$ from $j(x,t) \equiv -e n(x,t) v(x,t)$. Its Fourier transform is
\begin{eqnarray} \label{eq:DS_convolution}
j(x,\omega) = -e n_0 v(x,\omega) - e \sum_{\omega_1} \delta n(x,\omega_1) v(x,\omega_2)
~.
\end{eqnarray}
The product in time domain becomes in fact a convolution over the frequency.
Using Eq.~(\ref{eq:DS_convolution}), Eq.~(\ref{eq:DS_euler_2}) becomes
\begin{widetext}
\begin{eqnarray} \label{eq:DS_euler_4}
-\frac{e}{m} f(\omega) E(x,\omega) &=& 
\left(-i\omega+\frac{1}{\tau} \right) v(x,\omega) 
+
\sum_{\omega_1}
 v(x,\omega_2)\partial_x  v (x,\omega_1)
\nonumber\\
&+& 
\sum_{\omega_1} \left[\frac{g(\omega,\omega_1) }{f(\omega_1) f(\omega_2)} -1\right]
\big[
v(x,\omega_2)\partial_x  v (x,\omega_1)
+ i \omega \frac{\delta n (x,\omega_1)}{n_0} v(x,\omega_2)
\big]
~.
\end{eqnarray}
\end{widetext}
In Eq.~(\ref{eq:DS_euler_4}) we added a phenomenological damping term, by replacing $-i\omega \to -i\omega+1/\tau$ in the prefactor of the first term of Eq.~(\ref{eq:DS_euler_2}). 
At the same time, replacing the current with the velocity field, the continuity equation becomes
\begin{eqnarray} \label{eq:DS_continuity_2}
&&
\partial_t U(x, t) + \partial_x \big[ U (x,t) v(x,t)\big] = 0 
~.
\end{eqnarray}
We now determine the rectified potential, subject to the asymmetric DS boundary conditions
\begin{eqnarray} \label{eq:DS_linear_BC}
&&
U(x=0,t) = U_0 + U_{\rm ext} \cos(\omega t)
~,
\nonumber\\
&&
v(x=L,t) = 0
~,
\end{eqnarray}
i.e.~for an external electric potential oscillating at frequency $\omega$ applied between source and gate, while the drain is left fluctuating and no current flows through it. 

The problem can be solved in two steps, as in the standard case of DS theory~\cite{dyakonov_ieee_1996a}. First, the potential and velocity field are written as
\begin{eqnarray} \label{eq:nonlin_ansatz}
&&
U(x,t) = U_0 + U_1 (x,t) + U_2(x,t)
~,
\nonumber\\
&&
v(x,t) = v_0 + v_1 (x,t) + v_2(x,t)
~,
\end{eqnarray}
where $U_0 = -e n_0/C$ and $v_0 = 0$ represent their equilibrium properties. Next, we determine the eigenmodes of the linear problem defined by plugging Eq.~(\ref{eq:nonlin_ansatz}) into Eqs.~(\ref{eq:DS_euler_4})-(\ref{eq:DS_continuity_2}) and neglecting all terms containing $U_2(x,t)$, $v_2(x,t)$, as well as products of $U_1(x,t)$ and $v_1(x,t)$. Hence $U_1(x,t)$ and $v_1(x,t)$ are used to solve the linear problem and determine its eigenmodes.

The last step consists in (i) isolating all nonlinear terms in $U_1(x,t)$ and $v_1(x,t)$ in Eqs.~(\ref{eq:DS_euler_4})-(\ref{eq:DS_continuity_2}), together with those linear in $U_2(x,t)$ and $v_2(x,t)$, and (ii) solve such equations for $U_2(x,t)$ and $v_2(x,t)$. We therefore obtain the potential and velocity field resulting from the nonlinear interaction between linear eigenmodes. These contain both terms oscillating at frequency $2\omega$ and a rectified part. We will focus on the latter one, which is singled out by integrating the nonlinear Eqs.~(\ref{eq:DS_euler_4})-(\ref{eq:DS_continuity_2}) over a period of oscillation of the external field, a procedure which effectively produces equations for time-averaged quantities.

\subsection{Eigenmodes of the linear problem}
We look for solutions of the form
\begin{eqnarray}
U_1(x,t) &=& {\tilde U}_1(x, \omega) e^{i\omega t} + {\tilde U}^\ast_1(x, \omega) e^{i\omega t}
~,
\end{eqnarray}
whose Fourier transform is 
\begin{eqnarray}\label{eq:U_1_FT}
U_1(x,{\tilde \omega}) =  {\tilde U}_1(x, \omega) \delta_{{\tilde \omega}, \omega} + {\tilde U}^\ast_1(x, \omega)  \delta_{{\tilde \omega}, -\omega}
~.
\end{eqnarray}
Similarly, for $v_1(x,{\tilde \omega})$,
\begin{eqnarray}\label{eq:v_1_FT}
v_1(x,{\tilde \omega}) =  {\tilde v_1}(x, \omega) \delta_{{\tilde \omega}, \omega} + {\tilde v}_1^\ast(x, \omega)  \delta_{{\tilde \omega}, -\omega}
~.
\end{eqnarray}
The linearized versions of Eqs.~(\ref{eq:DS_euler_4})-(\ref{eq:DS_continuity_2}) for each of the two components on the right-hand sides of Eqs.~(\ref{eq:U_1_FT})-(\ref{eq:v_1_FT}) read
\begin{eqnarray} \label{eq:DS_linear_def}
&&
- i \omega {\tilde U}_1(x, \omega) + U_0 \partial_x {\tilde v}_1(x,\omega) = 0 
~,
\nonumber\\
&&
\frac{e}{m} f(\omega) \partial_x {\tilde U}_1({\bm r},\omega) =  \left(-i\omega +\frac{1}{\tau}\right) {\tilde v}_1(x, \omega)
~.
\end{eqnarray}
The solution of these equation with the DS boundary conditions (\ref{eq:DS_linear_BC}) reads as following:
\begin{eqnarray} \label{eq:DS_1D_linear_solution}
{\tilde U}_1(x, \omega) = \frac{U_{\rm ext}}{2} \left( \frac{e^{i \kappa x}}{1 + e^{2 i \kappa L}} + \frac{e^{-i \kappa x}}{1 + e^{-2 i \kappa L}} \right)
~.
\end{eqnarray}
Note that the function inside the round brackets goes to one at $x=0$. ${\tilde U}_1(x, \omega)$ therefore satisfies the DS boundary condition for the potential given in the first line of Eq.~(\ref{eq:DS_linear_BC}).
In Eq.~(\ref{eq:DS_1D_linear_solution}),
\begin{eqnarray}\label{eq:kappa}
\kappa = \frac{\omega}{s} \sqrt{\left(1+\frac{i}{\omega \tau}\right)f^{-1}(\omega)}
~,
\end{eqnarray}
and $s = \sqrt{-e U_0/m}$. Note that the plasmon wavevector is logarithmically suppressed for $\omega\to 2 E_{\rm F}$.
From Eq.~(\ref{eq:DS_1D_linear_solution}), the velocity field is readily determined as
\begin{eqnarray}
{\tilde v}_1(x, \omega) = \frac{\omega}{\kappa} \frac{U_{\rm ext}}{2 U_0} \left( \frac{e^{i \kappa x}}{1 + e^{2 i \kappa L}} - \frac{e^{-i \kappa x}}{1 + e^{-2 i \kappa L}} \right)
~.
\end{eqnarray}
Note that the function inside the round brackets goes to zero at $x=L$. ${\tilde v}_1(x, \omega)$ therefore satisfies the DS boundary condition for the velocity given in the second line of Eq.~(\ref{eq:DS_linear_BC}).

\subsection{The rectified potential}
We start again from Eqs.~(\ref{eq:DS_euler_4}) and~(\ref{eq:DS_continuity_2}), and we now isolate their nonlinear terms. Eq.~(\ref{eq:DS_euler_4}) gives
\begin{eqnarray} \label{eq:DS_continuity_nonlinear}
&&
\partial_t U_2(x, t) + \partial_x \big[ U_0 v_2(x,t)+U_1(x,t) v_1(x,t)\big] = 0 
~,
\end{eqnarray}
whose Fourier transform is
\begin{equation} \label{eq:DS_continuity_nonlinear_FT}
i {\tilde \omega} U_2(x, {\tilde \omega}) = \partial_x \Big[ U_0 v_2(x,{\tilde \omega}) + \sum_{\omega_1} U_1(x,{\tilde \omega}-\omega_1) v_1(x,\omega_1)\Big]
~.
\end{equation}
Since we are interested in the rectified nonlinear quantities, we consider equation for quantities averaged over a period $2\pi/\omega$ of oscillation of the external field. This corresponds to evaluate Eq.~(\ref{eq:DS_continuity_nonlinear_FT})  at ${\tilde \omega} = 0$. Using Eqs.~(\ref{eq:U_1_FT})-(\ref{eq:v_1_FT}), the convolution on the right-hand side of Eq.~(\ref{eq:DS_continuity_nonlinear_FT}) is readily evaluated. We get
\begin{equation} \label{eq:DS_continuity_nonlinear_FT_2}
\partial_x \big[ U_0 v_2(x,0) + {\tilde U}_1^\ast(x,\omega) {\tilde v}_1(x,\omega) + {\tilde U}_1(x,\omega) {\tilde v}_1^\ast(x,\omega) \big] = 0 
~.
\end{equation}
Solving Eq.~(\ref{eq:DS_continuity_nonlinear_FT_2}) with the DS boundary conditions (\ref{eq:DS_linear_BC}) on the velocity at $x=L$, which implies ${\tilde v}_1(L,\omega) = {\tilde v}_1^\ast(L,\omega) = v_2(L,0) = 0$, we get
\begin{equation} \label{eq:DS_continuity_nonlinear_FT_3}
v_2(x,0) = -\frac{ {\tilde U}_1^\ast(x,\omega) {\tilde v}_1(x,\omega) + {\tilde U}_1(x,\omega) {\tilde v}_1^\ast(x,\omega) }{U_0}
~.
\end{equation}

Following similar steps, the nonlinear part of Eq.~(\ref{eq:DS_continuity_2}) becomes
\begin{eqnarray} \label{eq:DS_nonlinear_euler_FT}
\frac{e}{m} \partial_x U_2(x,0) &=& 
\frac{v_2(x,0)}{\tau}
+
\frac{g(0,\omega) }{f(\omega) f(-\omega)}
\nonumber\\
&\times&
\big[ v(x,\omega)\partial_x  v^\ast (x,\omega) + v^\ast(x,\omega)\partial_x  v (x,\omega) \big]
~.
\nonumber\\
\end{eqnarray}
%
\begin{figure}[t]
\begin{center}
\begin{tabular}{c}
\begin{overpic}[width=1.0\columnwidth]{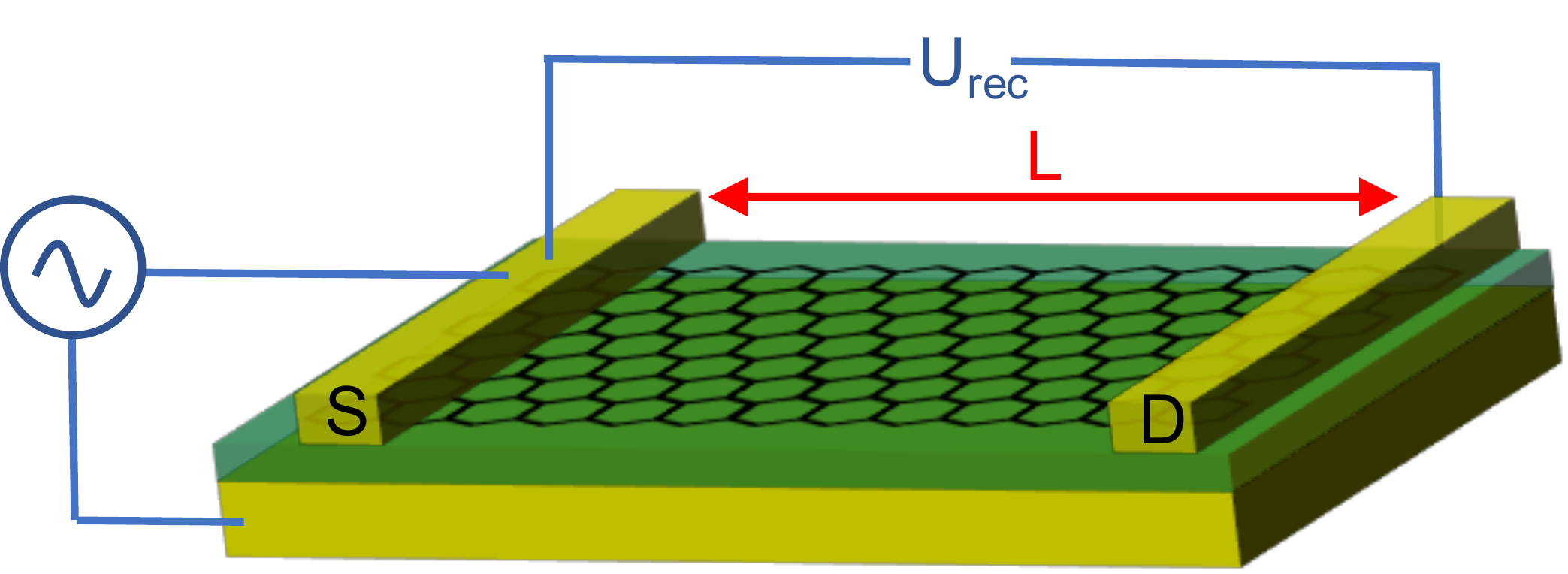}
\put(2,80){(a)}
\end{overpic}
\\
\begin{overpic}[width=1.0\columnwidth]{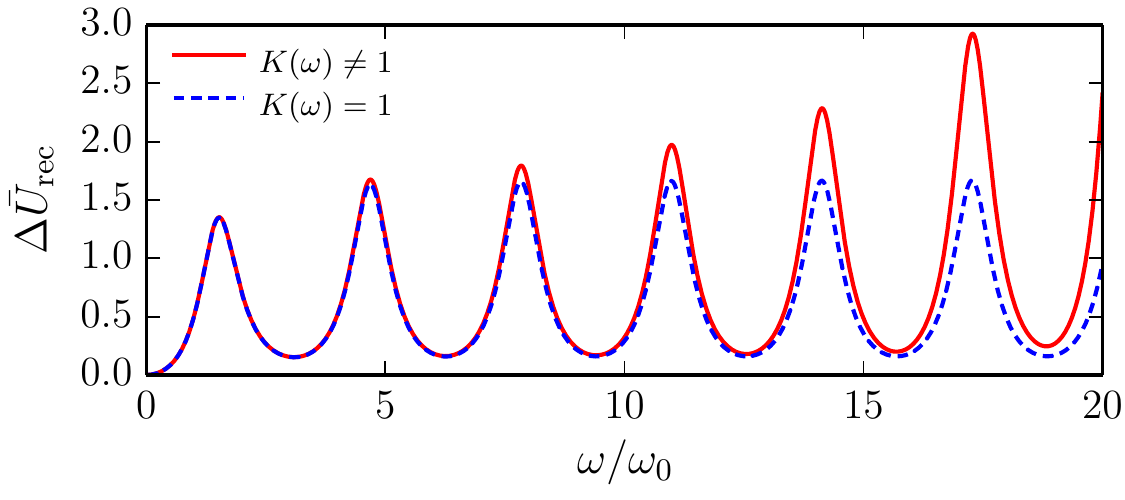}
\put(2,10){(b)}
\end{overpic}
\end{tabular}
\end{center}
\caption{(Color online)
Panel (a): the schematics of the 1D setup. A layer of graphene is placed on a substrate in close proximity to a metal gate and contacted at its ends. This defines a channel of length $L$ homogeneous in the transverse direction. The potential between the source and the bottom gate oscillates at the frequency $\omega$ with amplitude $U_{\rm ext}$. The drain is left fluctuating, {\it i.e.} no current flows though it. 
Panel (b): the dimensionless rectified potential $\Delta {\bar U}_{\rm rec} \equiv U_{0}\Delta U_{2}/U^2_{\rm ext}$ given by Eq.~(\ref{eq:rect_potential_result}) is plotted (solid line) as a function of the rescaled frequency $\omega/\omega_{0}$, where $\omega_0 = s/L$. For comparison, the standard DS result, obtained by setting $K(\omega)=1$ in Eq.~(\ref{eq:rect_potential_result}), corresponds to the the dashed curve. Numerical results in this plot have been obtained by setting $E_{\rm F} = 12 \omega_{0}$ and $\tau = \omega_0^{-1}$.
\label{fig:one}
}
\end{figure}
Note that the last term on the second line of Eq.~(\ref{eq:DS_euler_4}) yields no contribution to Eq.~(\ref{eq:DS_nonlinear_euler_FT}). There, in fact, $\omega \to {\tilde \omega}$ [using the same notation of Eq.~(\ref{eq:DS_continuity_nonlinear_FT})], and therefore such term vanishes in the limit ${\tilde \omega} \to 0$. 
We note that the rectified potential is defined as 
\begin{eqnarray}
\Delta U_2 = \int_0^L dx~\partial_x U_2(x,0)
~,
\end{eqnarray}
which allows to determine it by integrating Eq.~(\ref{eq:DS_nonlinear_euler_FT}), after having replaced the first term on the right-hand side with Eq.~(\ref{eq:DS_continuity_nonlinear_FT_3}).
Defining
\begin{equation}
K(\omega) \equiv \frac{g(0,\omega)}{f(\omega) f(-\omega)}~,
\end{equation}
$\kappa_1 = \Re e(\kappa)$, and $\kappa_2 = \Im m(\kappa)$, we finally find
\begin{eqnarray} \label{eq:rect_potential_result}
&&
\Delta U_2 = \frac{U_{\rm ext}^2}{4 U_0} \Bigg\{
1 - \frac{2}{\cos(2\kappa_1 L) + \cosh(2\kappa_2 L)} 
\nonumber\\
&&
+ \big[1 + K(\omega)\big] \frac{\omega \tau}{\sqrt{1+(\omega\tau)^2}} \frac{\cosh(2\kappa_2 L) - \cos(2\kappa_1 L)}{\cos(2\kappa_1 L) + \cosh(2\kappa_2 L)} 
\Bigg\}
~.
\nonumber\\
\end{eqnarray}
We remind the reader that the functions $f(\omega)$ and $g(\omega,\omega_{1})$ are defined in Eqs.~(\ref{eq:linear_conductivity_def}) and~(\ref{eq:g_function}), respectively. Eq.~(\ref{eq:rect_potential_result}) is the most important result of this Section on photodetection in 1D geometries with DS boundary conditions. The standard result of DS theory~\cite{dyakonov_ieee_1996a} is recovered by setting $K(\omega)=1$. In our pseudo-Euler theory this value is achieved only in the low-frequency ($\omega\ll 2 E_{\rm F}$) regime, as $\lim_{\omega\to 0}K(\omega) = 1$. At large frequencies, $\omega\sim 2E_{\rm F}$, however, we find a strong enhancement of the rectified signal as compared to DS theory, since $\lim_{\omega\to 2 E_{\rm F}}K(\omega) = +\infty$. 

As discussed after Eq.~(\ref{eq:d2xyyxM_main}), the enhancement of the nonlinear conductivity, which is at origin of the increased rectified signal for $\omega\to 2E_{\rm F}$, is a robust physical feature of the theory. We therefore expect it to be observable in a well-defined temperature window. A determination of such interval would require the inclusion of thermal broadening in the expressions for the nonlinear conductivity, which is beyond the scope of the present paper. We however observe that thermal broadening remains negligible (and therefore singularities well defined) as long as the temperature is small compared to the Fermi temperature. In graphene, for typical doping concentrations, the latter is of the order of $\sim 1000~{\rm K}$. Therefore, the enhancement of the rectified signal we predict could extend very well up to room temperature.

A plot of the rectified potential $\Delta U_2$ (solid line), in units of $U^2_{\rm ext}/U_{0}$, is given in Fig.~\ref{fig:one} as a function of the frequency $\omega$ of the external drive (measured in units of $\omega_0 = s/L$). The result is compared with the usual DS one~\cite{dyakonov_ieee_1996a} (dashed line), which, once again, is obtained by setting $K(\omega)=1$ in Eq.~(\ref{eq:rect_potential_result}).

\section{Theory of photodetection in 2D geometries: the low-frequency limit}
\label{sect:phot_2D}
We now switch to the case of a 2D graphene photodetector. For the sake of simplicity, we focus only in the low-frequency $\omega \ll 2E_{\rm F}$ regime.
We start again from Eq.~(\ref{eq:euler_non_hydro_2}), where now $\sigma(\omega) \to i n_0 e^2/(m \omega)$, while 
\begin{eqnarray}
d^{({\rm M})}_{xxyy}(\omega,\omega_1) &\to& -\frac{2 d_0}{\omega \omega_1 \omega_2}
~,
\nonumber\\
d^{({\rm Q})}_{xxyy} (\omega, \omega_1) &\to& - \frac{d_0}{\omega\omega_1^2}
~,
\nonumber\\
d^{({\rm Q})}_{xyxy} (\omega, \omega_1) &\to& \frac{d_0 (2\omega_1 + 3\omega_2)}{\omega \omega_1^2\omega_2}
~.
\end{eqnarray}
After few straightforward algebraic manipulations we get
\begin{widetext}
\begin{eqnarray} \label{eq:euler_non_hydro_4}
\frac{n_0 e^2}{m} {\bm E}(\Omega) &=& 
-i\omega {\bm j}(\Omega) - \frac{1}{n_0 e} \sum_{\Omega_1} \Big\{i {\bm q}_1 \big[{\bm j}(\Omega_1)\cdot {\bm j}(\Omega_2)\big] - i \omega_1 e \delta n(\Omega_1) {\bm j}(\Omega_2) 
-
i \omega_2 e \delta n(\Omega_1) {\bm j}(\Omega_2) \Big\}
\nonumber\\
&-&
\frac{1}{n_0 e} \sum_{\Omega_1} \Bigg\{ \left(\frac{\omega_2}{\omega_1} - 1 \right)
\big[i{\bm q}_1\times {\bm j} (\Omega_1)\big] \times {\bm j} (\Omega_2)
-
\left(1 + 2 \frac{\omega_2}{\omega_1}\right)
(i{\bm q}_1) \times \big[ {\bm j} (\Omega_1) \times {\bm j} (\Omega_2)\big] 
\Bigg\}
~.
\end{eqnarray}
\end{widetext}
In this equation, we separated the non-hydrodynamic terms stemming from the microscopic form of $\sigma^{(2)}_{ijk}(\Omega,\Omega_1,\Omega_2)$ from those that would appear in the canonical Euler equation, which are collected in the first line. This makes clear that, if ${\bm j} (\Omega_1)$ and ${\bm j} (\Omega_2)$ are longitudinal currents that propagate in the same direction (as in the 1D photodetection setup discussed in Sect.~\ref{sect:photo_1D}), all terms on the second line of Eq.~(\ref{eq:euler_non_hydro_4}) disappear. We conclude that, in the low-frequency regime, there is no difference between our pseudo-Euler theory and an approach {\it \`a la} DS based on the standard hydrodynamic Euler equation. We now show that, even in the low-frequency regime, qualitative differences emerge when one retains the last line of Eq.~(\ref{eq:euler_non_hydro_4}).

\begin{figure}[t]
\begin{center}
\begin{tabular}{c}
\begin{overpic}[width=1.0\columnwidth]{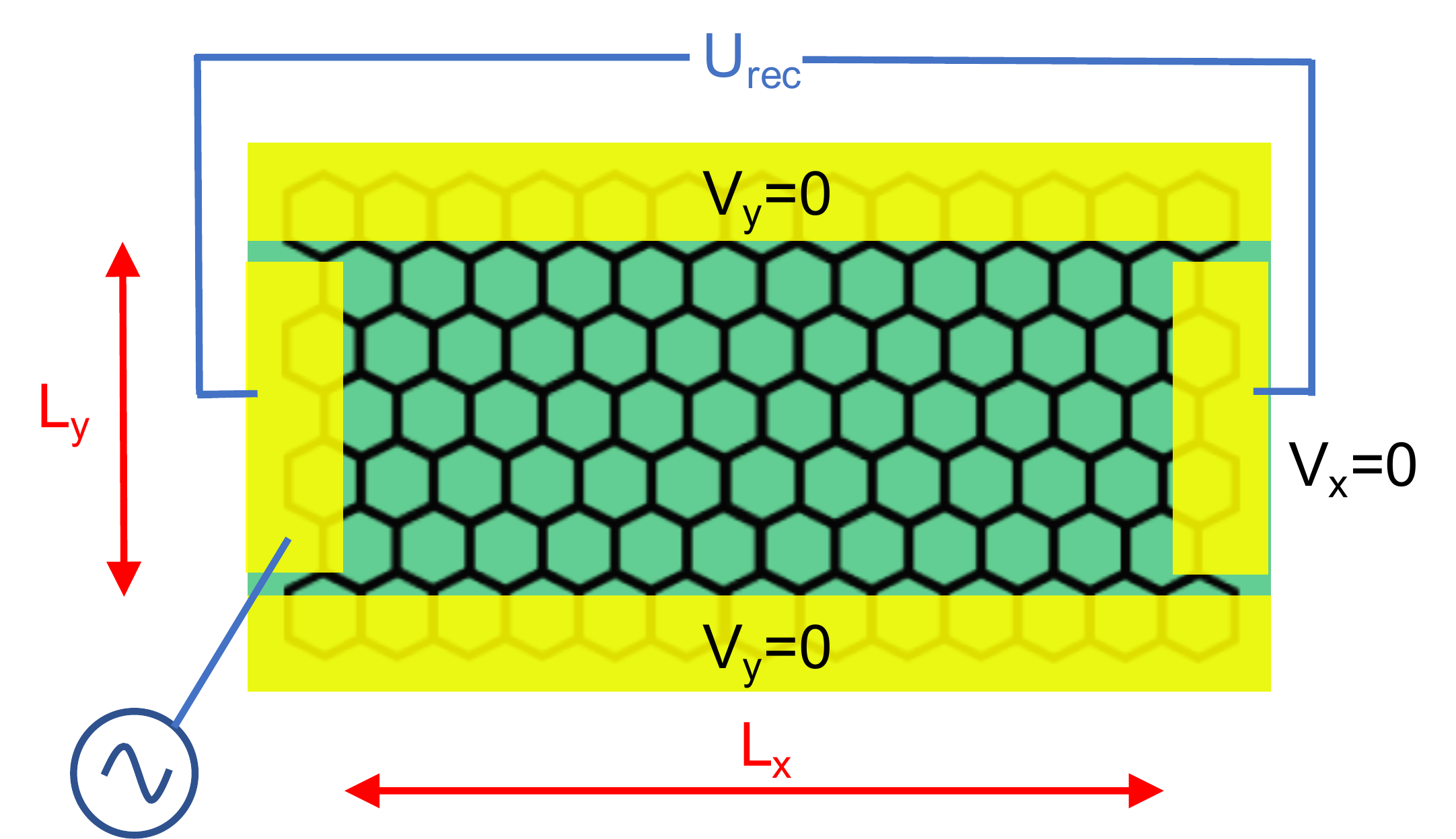}
\put(2,120){(a)}
\end{overpic}
\\
\begin{overpic}[width=1.0\columnwidth]{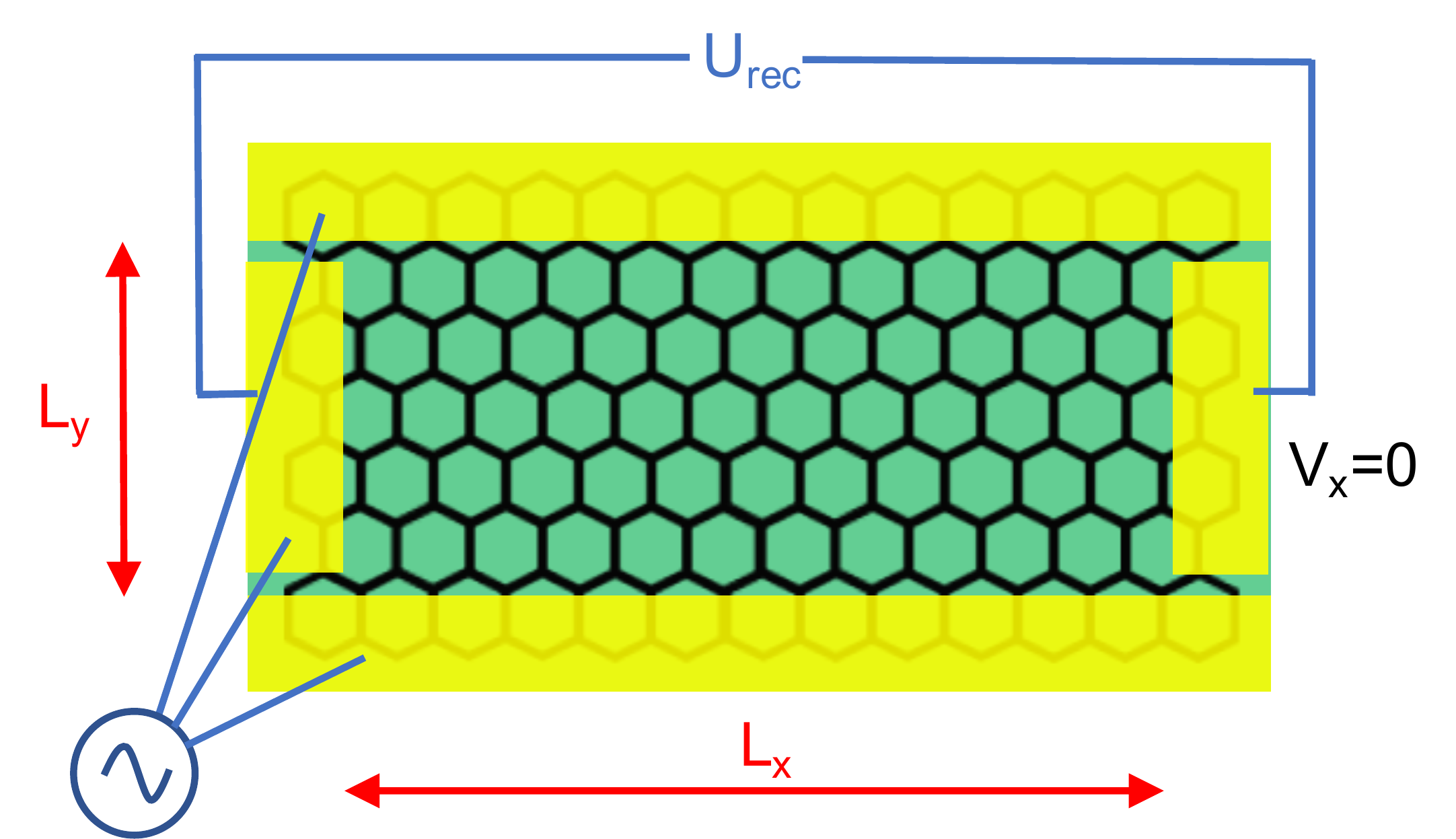}
\put(2,120){(b)}
\end{overpic}
\end{tabular}
\end{center}
\caption{(Color online)
Schematics of the 2D setup. A layer of graphene is placed on a substrate in close proximity to a gate (not shown) and contacted on all four sides. This defines a channel of length $L_x$ and width $L_y$. The rightmost contact is left fluctuating in both cases, whereas the leftmost oscillates at the frequency $\omega$. The rectified potential is always measured between the contacts at $x=0$ and $x=L_x$. The other two contacts ($y=0$, $y=L_y$) are either left fluctuating [panel (a)] or they are oscillating at the external frequency $\omega$ [panel (b)]. These two configurations correspond to the two boundary conditions studied in the main text in Sect.~\ref{sect:phot_2D}.
\label{fig:two}
}
\end{figure}

To this end, we manipulate the last line of Eq.~(\ref{eq:euler_non_hydro_4}) as following. We define ${\bm \alpha}({\bm q},\omega)$ such that
\begin{eqnarray}
{\bm j}({\bm q},\omega) = - i \omega (-e n_0) {\bm \alpha}({\bm q},\omega)
~,
\end{eqnarray}
which in the real-time domain implies that
\begin{eqnarray}
{\bm j}({\bm r},t) = -e n_0 \partial_t {\bm \alpha}({\bm r},t)
~.
\end{eqnarray}
Fourier transforming to real space and time Eq.~(\ref{eq:euler_non_hydro_4}), we get
\begin{widetext}
\begin{eqnarray} \label{eq:euler_non_hydro_5}
\frac{n_0 e^2}{m} {\bm E}({\bm r},t) &=& 
\partial_t {\bm j}({\bm r}, t) 
- \frac{1}{n_0 e} \Big\{\sum_i j_i({\bm r}, t) {\bm \nabla} j_i({\bm r}, t) + e \partial_t \delta n({\bm r},t) {\bm j}({\bm r}, t) 
+e \delta n({\bm r},t) \frac{n_0 e^2}{m} {\bm E}({\bm q},\omega)\Big\}
\nonumber\\
&+&
\frac{1}{n_0 e} 
\Big\{
\big[{\bm \nabla} \times {\bm j} ({\bm r},t)\big] \times {\bm j} ({\bm r},t)
+
\big[{\bm j} ({\bm r},t) \cdot{\bm \nabla}\big] {\bm j} ({\bm r},t) 
-
\big[{\bm \nabla}\cdot{\bm j} ({\bm r},t)\big] {\bm j} ({\bm r},t) 
\Big\}
\nonumber\\
&+&
\frac{n_0 e^2}{m} \Big\{ 
\big[{\bm \nabla} \times {\bm \alpha}({\bm r},t)\big] \times  {\bm E} ({\bm r},t)
-
2
\big[{\bm E} ({\bm r},t) \cdot{\bm \nabla}\big] {\bm \alpha} ({\bm r},t) 
+
2 \big[{\bm \nabla}\cdot{\bm \alpha} ({\bm r},t)\big] {\bm E} ({\bm r},t) 
\big] 
\Big\}
~.
\end{eqnarray}
In this equation we used that, to the order we are working, in the nonlinear terms we can replace $\partial_t {\bm j}({\bm r}, t) = n_0 e^2 {\bm E}({\bm r},t)/m$.
We now recast it in an Euler-like form. As usual, we define
${\bm j}({\bm r},t) = - e n({\bm r}, t) {\bm v}({\bm r},t)$
where clearly at linear order
${\bm v}({\bm r},t) = \partial_t {\bm \alpha}({\bm r},t)$.
Dividing by $-e n({\bm r}, t)$, and neglecting higher-order nonlinear terms, Eq.~(\ref{eq:euler_non_hydro_5}) finally becomes
\begin{eqnarray} \label{eq:euler_non_hydro_8_rew}
-\frac{e}{m} {\bm E}({\bm r},t) &=& 
\partial_t {\bm v}({\bm r}, t) + \frac{1}{\tau} {\bm v}({\bm r}, t) + \frac{1}{2} {\bm \nabla} v^2({\bm r},t)
-\lambda
\Big\{
2\big[{\bm \nabla} \times {\bm v} ({\bm r},t)\big] \times {\bm v} ({\bm r},t)
+
\big[{\bm v} ({\bm r},t) \times {\bm \nabla}\big] \times {\bm v} ({\bm r},t)
\Big\}
\nonumber\\
&+&
\frac{e\lambda}{m} \Big\{ 
\big[{\bm \nabla} \times {\bm \alpha}({\bm r},t)\big] \times  {\bm E} ({\bm r},t)
+
2 \big[{\bm E} ({\bm r},t) \times {\bm \nabla}\big] \times {\bm \alpha}({\bm r},t)
\Big\}
~.
\end{eqnarray}
\end{widetext}
This is the main result of this Section. The first four terms on the first line of Eq.~(\ref{eq:euler_non_hydro_8_rew}) reproduce the canonical Euler equation, with the addition of the phenomenological damping term ${\bm v}({\bm r}, t)/\tau$. All other terms are entirely new and stem from the microscopic nonlinear optical properties of graphene. Note that we introduced the book-keeping parameter $\lambda$, which allows to interpolate between the hydrodynamic ($\lambda=0$) and non-hydrodynamic ($\lambda=1$) regimes.
Within the local-capacitance approximation, Eq.~(\ref{eq:euler_non_hydro_8_rew}) is to be solved together with
\begin{eqnarray} \label{eq:cont_NS_poisson_rew}
\partial_t U({\bm r}, t) + {\bm \nabla}\cdot \big[U({\bm r},t){\bm v}({\bm r},t)\big] = 0 
~,
\end{eqnarray}
as well as appropriate boundary conditions. Concrete numerical results will be obtained for a rectangular sample with sides of length $L_x$ and $L_y$. We will show results for two cases:
\begin{itemize}
\item[a)] An electrode, placed at the side $x=0$, oscillates at frequency $\omega$, while all other sides of the rectangle are left fluctuating. Hence, no current flows through them and
\begin{eqnarray} \label{eq:linear_BC_case_1}
U(0,y,t) &=& U_0 + U_{\rm ext} \cos(\omega t)
~,
\nonumber\\
v_x(L_x,y,t) &=& v_y(x,0,t) = v_y(x,L_y,t) = 0
~.
\end{eqnarray}
See also Fig.~\ref{fig:two}(a).
This case reproduces the standard DS results and serves as a test of the procedure, as well as of the numerical implementation. 

\item[b)] Three electrodes, at the sides $x = 0$, $y = 0$ and $y = L_y$, oscillate with frequency $\omega$, whereas the last one is left fluctuating. Hence,
\begin{eqnarray} \label{eq:linear_BC_case_2}
U(0,y,t) &=& U(x,0,t) = U(x,L_y,t) 
\nonumber\\
&=& U_0 + U_{\rm ext} \cos(\omega t)~,
\nonumber\\
v_x(L_x,y,t) &=& 0
~.
\end{eqnarray}
See also Fig.~\ref{fig:two}(b).
In this truly 2D case we will be able to appreciate qualitative differences between the standard hydrodynamic approach ($\lambda=0$) and our pseudo-Euler theory ($\lambda=1$). 
\end{itemize}
For the sake of simplicity, numerics will be carried out for $L_x = L_y = L$.

\subsection{Eigenmodes of the linear problem}
When we linearize Eqs.~(\ref{eq:euler_non_hydro_8_rew})-(\ref{eq:cont_NS_poisson_rew}) we get
\begin{eqnarray} \label{eq:cont_NS_poisson_linear}
&&
\partial_t U_1({\bm r}, t) + U_0 {\bm \nabla}\cdot {\bm v}_1({\bm r},t) = 0 
~,
\nonumber\\
&&
\partial_t {\bm v}_1({\bm r}, t) + \frac{1}{\tau} {\bm v}_1({\bm r}, t) = \frac{e}{m} {\bm \nabla} U_1({\bm r},t)
~,
\end{eqnarray}
which results in the equation
\begin{eqnarray} \label{eq:cont_NS_poisson_linear_2}
&&
\partial_t^2 U_1({\bm r}, t) + \frac{1}{\tau} \partial_t U_1({\bm r}, t) - s^2 {\bm \nabla}^2 U_1({\bm r}, t) = 0 
~.
\end{eqnarray}
The frequency $\omega$ is fixed, so we attempt a solution of the type
\begin{eqnarray}
U_1({\bm r}, t) = U_1({\bm r},\omega) e^{-i\omega t} + U_1^\ast({\bm r},\omega) e^{i\omega t}
~,
\end{eqnarray}
which gives
\begin{eqnarray} \label{eq:cont_NS_poisson_linear_3}
&&
{\bm \nabla}^2 U_1({\bm r},\omega) = -q_\omega^2 U_1({\bm r},\omega)
~,
\nonumber\\
&&
{\bm \nabla}^2 U_1^\ast({\bm r},\omega) = -(q^\ast_\omega)^2 U_1^\ast({\bm r},\omega)
~,
\end{eqnarray}
where 
\begin{eqnarray}
&&
q _\omega = \frac{\omega}{s} \sqrt{1 + \frac{i}{\omega \tau}}
~,
\nonumber\\
&&
q _\omega^\ast = \frac{\omega}{s} \sqrt{1 - \frac{i}{\omega \tau}}
~.
\end{eqnarray}
The harmonic problem posed by Eqs.~(\ref{eq:cont_NS_poisson_linear_3}) can be solved analytically or numerically with the appropriate boundary conditions at~(\ref{eq:linear_BC_case_1}) or~(\ref{eq:linear_BC_case_2}). Eq.~(\ref{eq:cont_NS_poisson_linear}) implies that, at the contacts oscillating at frequency $\omega$, 
\begin{eqnarray}
U_1({\bm r},\omega)\Big|_{\rm bound} = \frac{U_{\rm ext}}{2}
~,
\end{eqnarray}
whereas at the fluctuating contacts
\begin{eqnarray}
\partial_{\hat {\bm n}} U_1({\bm r},\omega)\Big|_{\rm bound} = 0
~,
\end{eqnarray}
where ``$\partial_{\hat {\bm n}}$'' denotes the gradient in the direction orthogonal to the boundary.

Once $U_1({\bm r},\omega)$ and $U_1^\ast({\bm r},\omega)$ have been determined, the velocity field is found as
\begin{eqnarray}  \label{eq:general_linear_solution_result}
{\bm v}_1({\bm r}, t) = {\bm v}_1({\bm r},\omega) e^{-i\omega t} + {\bm v}_1^\ast({\bm r},\omega) e^{i\omega t}
~,
\end{eqnarray}
where
\begin{eqnarray} \label{eq:2D_linear_v1_sol}
&&
{\bm v}_1({\bm r},\omega) = \frac{e {\bm \nabla}U_1({\bm r},\omega)/m}{- i \omega + 1/\tau}
~,
\nonumber\\
&&
{\bm v}_1^\ast({\bm r},\omega) = \frac{e {\bm \nabla}U_1^\ast({\bm r},\omega)/m}{i \omega + 1/\tau}
~,
\end{eqnarray}
where ${\bm \alpha}_1({\bm r},\omega) = i {\bm v}_1({\bm r},\omega)/\omega$ and ${\bm \alpha}_1^\ast({\bm r},\omega) = -i {\bm v}_1^\ast({\bm r},\omega)/\omega$.

\begin{figure}[t]
\begin{center}
\begin{tabular}{c}
\begin{overpic}[width=1.0\columnwidth]{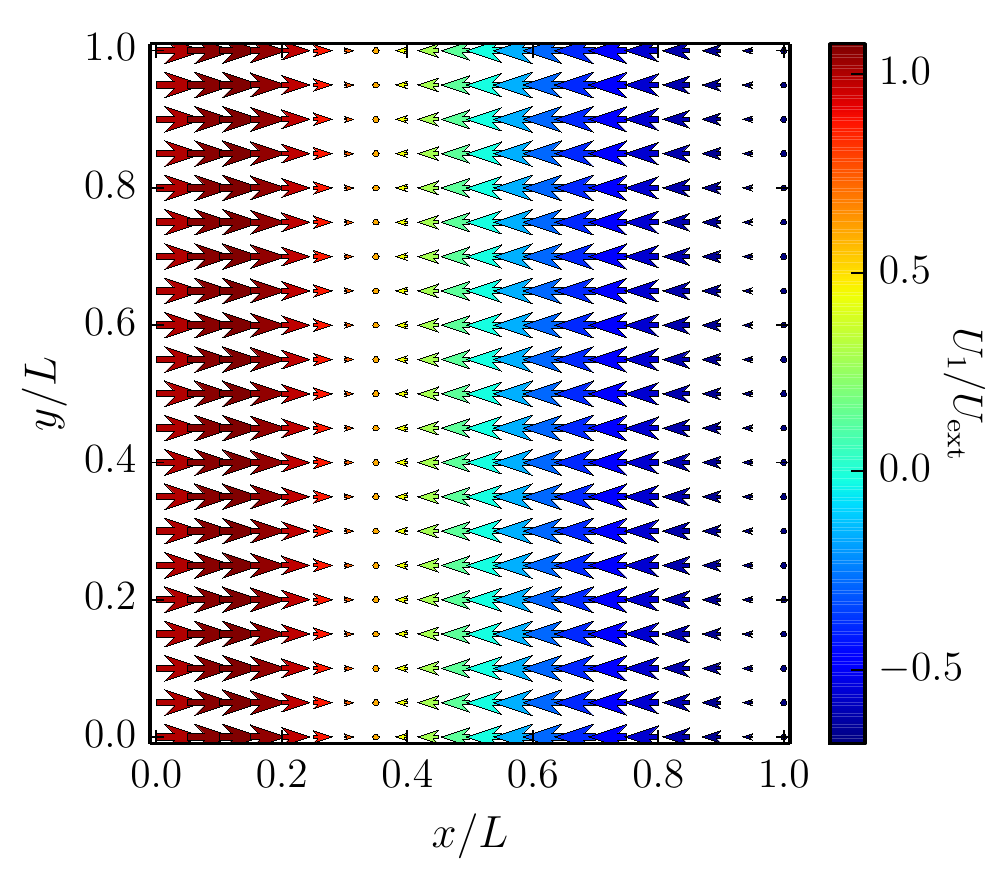}\put(2,200){(a)}
\end{overpic}
\\
\begin{overpic}[width=1.0\columnwidth]{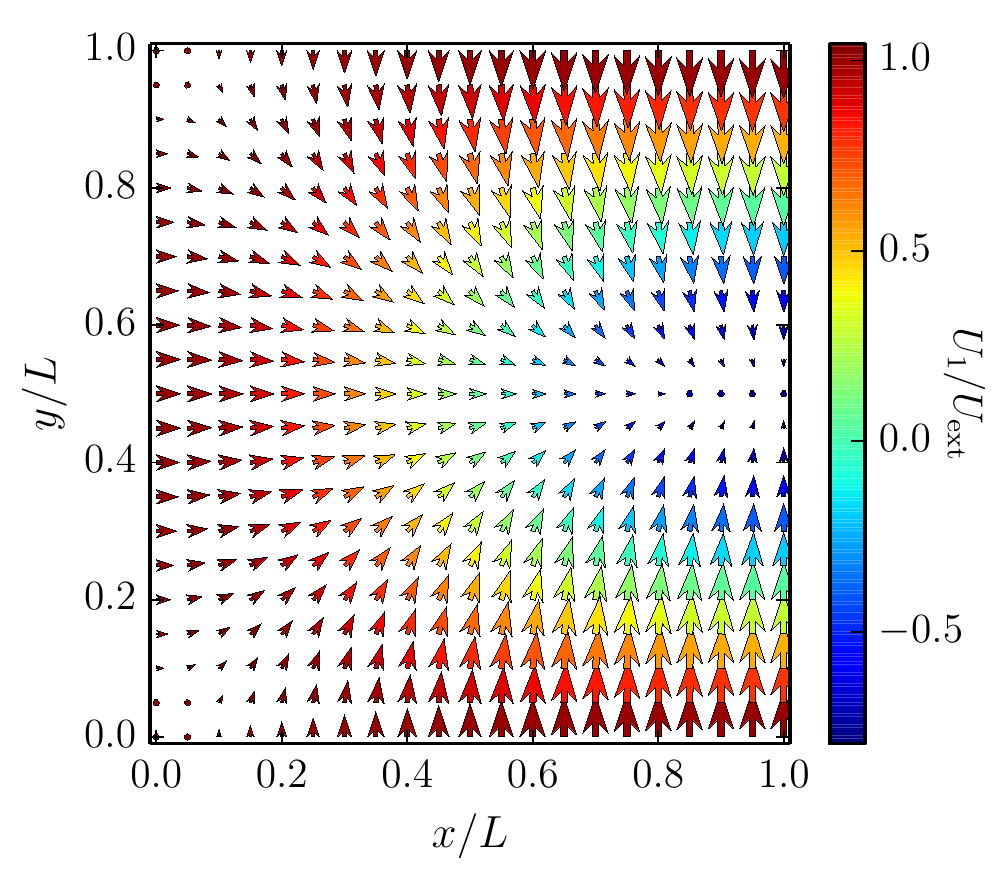}\put(2,200){(b)}
\end{overpic}
\end{tabular}
\end{center}
\caption{(Color online) 
\label{fig:three}
The velocity profile ${\bm v}_1({\bm r},0)$ for the boundary conditions in Eq.~(\ref{eq:linear_BC_case_1}) [panel (a)] and~(\ref{eq:linear_BC_case_2}) [panel (b)], in units of $s U_{\rm ext}/U_0$. The color of the arrow corresponds to the value of $U_1({\bm r},0)/U_{\rm ext}$. In panel (a), $\omega = 4.5\omega_0$, whereas in panel (b) $\omega = 3.5\omega_0$. Numerical results in both panels have been obtained by setting $\tau = \omega^{-1}_0$ and $L_x = L_y = L$. (Due to the chosen scalings, these are the only quantities that matter for making the plots.) As can be seen from Fig.~\ref{fig:four}, the frequencies of both plots are chosen to be close to a resonant frequency of the purely hydrodynamic FET ({\it i.e.} when $\lambda=0$).
}
\end{figure}

In Fig.~\ref{fig:three} we show representative velocity fields, ${\bm v}_{1}({\bm r},0)$ for both cases defined by the boundary conditions in Eqs.~(\ref{eq:linear_BC_case_1}) and~(\ref{eq:linear_BC_case_2}). In the former case the problem is effectively 1D, since velocities and potentials are all homogeneous in the ${\hat {\bm y}}$ direction. 
Note that, in this plot, we need not to distinguish between hydrodynamic and non-hydrodynamic regimes: the extra terms only impact the nonlinear part of the problem and the final rectified potential.

\subsection{The rectified potential}
Once the solution of the harmonic problem posed by Eq.~(\ref{eq:cont_NS_poisson_linear_3}), with the boundary conditions~(\ref{eq:linear_BC_case_1}) or~(\ref{eq:linear_BC_case_2}), has been analytically or numerically determined, we can solve the nonlinear problem. We now guide the reader through the steps of the solution.

First, we collect all the nonlinear terms in Eqs.~(\ref{eq:cont_NS_poisson_rew}) and~(\ref{eq:euler_non_hydro_8_rew}), which give the equations
\begin{equation} \label{eq:cont_nonlinear}
\partial_t U_2({\bm r}, t) + {\bm \nabla}\cdot \big[U_0 {\bm v}_2({\bm r},t) + U_1({\bm r},t){\bm v}_1({\bm r},t)\big] = 0
\end{equation}
and
\begin{eqnarray} \label{eq:NS_nonlinear}
\frac{e}{m} {\bm \nabla} U_2({\bm r},t) &=&
\partial_t {\bm v}_2({\bm r}, t) + \frac{1}{\tau} {\bm v}_2({\bm r}, t) + \frac{1}{2} {\bm \nabla} v_1^2({\bm r},t)
\nonumber\\
&-&
\lambda
\big[{\bm v}_1 ({\bm r},t) \times {\bm \nabla}\big] \times {\bm v}_1 ({\bm r},t)
\nonumber\\
&-&
\frac{2 e\lambda}{m}
\big[{\bm \nabla}U_1 ({\bm r},t) \times {\bm \nabla}\big] \times {\bm \alpha}_1({\bm r},t)
~.
\end{eqnarray}
Here, we used the solution of the linear problem, which, due to Eq.~(\ref{eq:2D_linear_v1_sol}), has the following properties: ${\bm \nabla} \times {\bm v}_1({\bm r}, t) = 0$ and ${\bm \nabla} \times {\bm \alpha}_1({\bm r}, t)$. Averaging Eqs.~(\ref{eq:cont_nonlinear})-(\ref{eq:NS_nonlinear}) over one period of the external oscillating field as
\begin{eqnarray}
\langle A ({\bm r}, t)\rangle = \frac{\omega}{2\pi}\int_0^{2\pi/\omega} dt~A({\bm r},t)
~,
\end{eqnarray}
we find the equations for the rectified quantities
\begin{equation} \label{eq:cont_nonlinear_2}
{\bm \nabla}\cdot \big[U_0 \langle {\bm v}_2 ({\bm r})\rangle + \langle U_1({\bm r},t){\bm v}_1({\bm r},t)\rangle \big] = 0
\end{equation}
and
\begin{eqnarray} \label{eq:cont_NS_nonlinear_2}
\frac{e}{m} {\bm \nabla} \langle U_2({\bm r},t)\rangle &=& 
\frac{1}{\tau} \langle{\bm v}_2({\bm r}, t) \rangle+ \frac{1}{2} {\bm \nabla} \langle v_1^2({\bm r},t)\rangle
\nonumber\\
&-&
\lambda
\big\langle
\big[{\bm v}_1 ({\bm r},t) \times {\bm \nabla}\big] \times {\bm v}_1 ({\bm r},t)
\big\rangle
\nonumber\\
&-& 
2\frac{e}{m} 
\big\langle
\big[{\bm \nabla}U_1 ({\bm r},t) \times {\bm \nabla}\big] \times {\bm \alpha}_1({\bm r},t)
\big\rangle
~.
\nonumber\\
\end{eqnarray}
As shown in Appendix~\ref{app:simplification_nonlinear}, Eqs.~(\ref{eq:cont_nonlinear_2}) and~(\ref{eq:cont_NS_nonlinear_2}) can be rewritten as
\begin{equation} \label{eq:cont_NS_nonlinear_3_1_main}
{\bm \nabla}\cdot \big[U_0 {\bm v}_2 ({\bm r}) + U_1^\ast({\bm r},\omega) {\bm v}_1({\bm r}, \omega) + U_1({\bm r},\omega) {\bm v}_1^\ast({\bm r}, \omega)   \big] = 0
\end{equation}
and
\begin{eqnarray} \label{eq:cont_NS_nonlinear_3_3_main}
&&
{\bm \nabla} \phi({\bm r}) =
\frac{2 i \lambda}{\omega \tau} {\bm \nabla}\times \big[ {\bm v}_1^\ast({\bm r}, \omega) \times {\bm v}_1({\bm r},\omega) \big]
\nonumber\\
&&
+
\frac{U_0 {\bm v}_2({\bm r}) + \lambda \big[ U_1^\ast({\bm r},\omega) {\bm v}_1({\bm r}, \omega) + U_1({\bm r},\omega) {\bm v}_1^\ast({\bm r}, \omega) \big] }{U_0 \tau}
~,
\nonumber\\
\end{eqnarray}
where
\begin{eqnarray} \label{eq:phi_def_main}
\phi({\bm r}) &\equiv& \frac{e}{m} U_2({\bm r}) - (1+\lambda){\bm v}_{1}({\bm r}, \omega) \cdot {\bm v}_{1}^\ast({\bm r},\omega) 
\nonumber\\
&-&
\lambda \frac{s^2}{U_0^2} U_1^\ast({\bm r},\omega) U_1({\bm r},\omega)
~,
\end{eqnarray}
$U_2({\bm r}) \equiv \langle U_2({\bm r},t)\rangle$, and ${\bm v}_2({\bm r}) \equiv \langle {\bm v}_2({\bm r},t)\rangle$. Eqs.~(\ref{eq:cont_NS_nonlinear_3_1_main})-(\ref{eq:cont_NS_nonlinear_3_3_main}) are written in a form that allows for an efficient and stable numerical evaluation.
It is in fact now possible to take the divergence of Eq.~(\ref{eq:cont_NS_nonlinear_3_3_main}) and then, by using Eq.~(\ref{eq:cont_NS_nonlinear_3_1_main}), we get
\begin{equation} \label{eq:cont_NS_nonlinear_final}
{\bm \nabla}^2 \phi({\bm r}) = 
\frac{\lambda - 1}{U_0 \tau} {\bm \nabla}\cdot\big[ U_1^\ast({\bm r},\omega) {\bm v}_1({\bm r}, \omega) + U_1({\bm r},\omega) {\bm v}_1^\ast({\bm r}, \omega) \big]
~.
\end{equation}
The boundary condition of zero velocity across any boundary can be translated, by using Eq.~(\ref{eq:cont_NS_nonlinear_3_3_main}), into a boundary condition on the normal derivative of $\phi({\bm r})$ as
\begin{equation} \label{eq:cont_NS_nonlinear_3_4}
\partial_{{\hat {\bm n}}} \phi({\bm r}) \Big|_{\rm bound} =
\frac{2 i \lambda}{\omega \tau} {\hat {\bm n}}\cdot \big\{{\bm \nabla}\times \big[ {\bm v}_1^\ast({\bm r}, \omega) \times {\bm v}_1({\bm r},\omega) \big]\big\}
~.
\end{equation}
On any other boundary, it is sufficient to impose that $U_2({\bm r}) = 0$, which according to Eq.~(\ref{eq:phi_def_main}) gives
\begin{eqnarray} 
\phi({\bm r}) \Big|_{\rm bound} &=& - (1+\lambda){\bm v}_{1}({\bm r}, \omega) \cdot {\bm v}_{1}^\ast({\bm r},\omega) 
\nonumber\\
&-&
\lambda \frac{s^2}{U_0^2} U_1^\ast({\bm r},\omega) U_1({\bm r},\omega)
~.
\end{eqnarray}
Once $\phi({\bm r})$ has been determined, the spatially-averaged rectified potential $\Delta U_{\rm rec}$ is obtained by inverting Eq.~(\ref{eq:phi_def_main}) for $U_2({\bm r})$ and then calculating
\begin{eqnarray}\label{eq:final_result_2D}
\Delta U_{\rm rec} \equiv \frac{1}{L_y} \int_0^{L_y} dy~\big[U_2(L_x,y) - U_2(0,y)\big]
~.
\end{eqnarray}
%

\begin{figure}[t]
\begin{center}
\begin{tabular}{c}
\begin{overpic}[width=1.0\columnwidth]{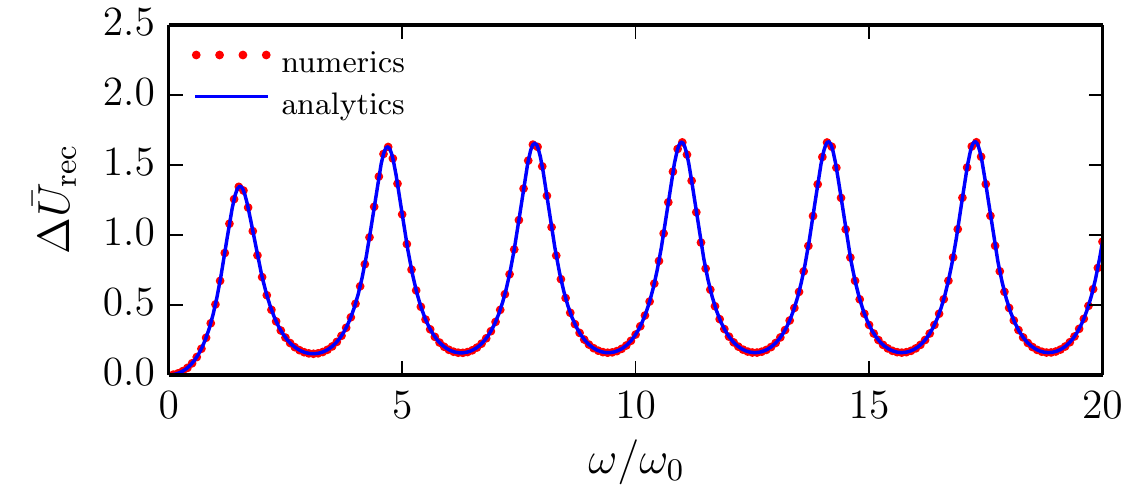}\put(2,110){(a)}
\end{overpic}
\\
\begin{overpic}[width=1.0\columnwidth]{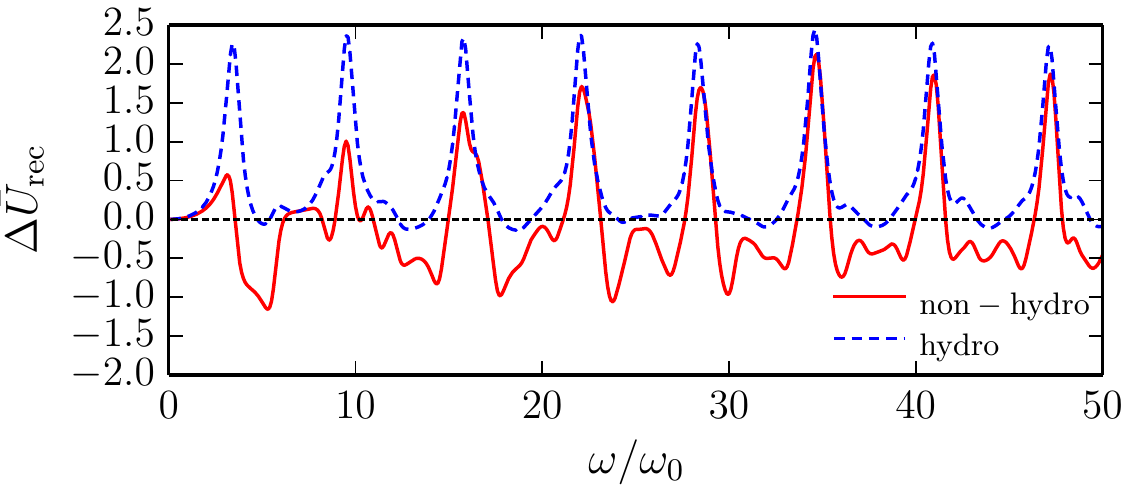}\put(2,110){(b)}
\end{overpic}
\end{tabular}
\end{center}
\caption{(Color online)
The spatially-averaged dimensionless rectified potential $\Delta {\bar U}_{\rm rec} \equiv U_0 \Delta U_{\rm rec}/U_{\rm ext}^2$ calculated from Eq.~(\ref{eq:final_result_2D}) is plotted as a function of the frequency of the external drive, in units of $\omega_0 = s/L$. Panel (a) [(b)] is obtained by imposing the boundary conditions in Eq.~(\ref{eq:linear_BC_case_1}) [(\ref{eq:linear_BC_case_2})].
In panel (a) we compare the numerical result to the analytical one. 
Panel (b) highlights the qualitative differences between hydrodynamic ($\lambda=0$) and non-hydrodynamic ($\lambda=1$) theories. Numerical results in this figure have been obtained by setting $\tau = \omega_0^{-1}$ and $L_x = L_y = L$.
\label{fig:four}
}
\end{figure}

In Fig.~\ref{fig:four}(a) we show a representative plot of $\Delta U_{\rm rec}$, in units of $U_{\rm ext}^2/U_0$, for the case obtained by imposing the boundary conditions~(\ref{eq:linear_BC_case_1}) for $\lambda =0, 1$ (as explained above, the two solutions coincide). The numerical result is compared with the analytical one, given by Eq.~(\ref{eq:rect_potential_result}) with $K(\omega)=1$. The excellent agreement shows that the numerical finite-element code we use, to solve the linear and nonlinear problems posed by Eqs.~(\ref{eq:cont_NS_poisson_linear_3}) and~(\ref{eq:cont_NS_nonlinear_final}) respectively, is stable and reliable.

In Fig.~\ref{fig:four}(b) we show instead a plot of $\Delta U_{\rm rec}$, always in units of $U_{\rm ext}^2/U_0$, for the case in which the boundary conditions are given by Eq.~(\ref{eq:linear_BC_case_2}), for both $\lambda=0$ and $\lambda=1$. In this case, as expected, the two results not only do not coincide but display large qualitative differences. In particular, note that the rectified potential calculated with our pseudo-Euler equation shows wild sign changes as a function of $\omega$. 

These qualitative differences between hydrodynamic and non-hydrodynamic regimes can potentially be probed experimentally by using antenna-coupled graphene THz photodetectors similar to those fabricated in Refs.~\onlinecite{vicarelli_naturemat_2012,spirito_apl_2014,bianco_apl_2015,bandurin_arxivTHz_2018}. To this end, one needs to minimize thermoelectric contributions~\cite{koppens_naturenano_2014,bandurin_apl_2018} to the dc photovoltage by making sure that the FET channel is as homogeneous as possible, mitigating the impact of gradients of the Seebeck coefficient. We further note, that while the detection of THz radiation enabled by conventional hydrodynamic nonlinearities remains to be observed, resonant (plasmon-assisted) THz photoresponse in graphene FETs has been recently reported at cryogenic temperatures~\cite{bandurin_arxivTHz_2018}, well below those required for the onset of hydrodynamic electron transport~\cite{bandurin_science_2016,bandurin_arxiv_2018}. Importantly, the resonances were well-pronounced even for the case of uniform channel doping, where the thermoelectric contribution to the responsivity is suppressed~\cite{bandurin_apl_2018}. This suggests that the pseudo-Euler nonlinearity addressed in detail in this Article may be one of the mechanisms responsible for the rectification of the high-frequency ac field into a dc photovoltage in Ref.~\onlinecite{bandurin_arxivTHz_2018}.

\section{Summary and conclusions}
\label{sect:summary}

In this Article we have revisited the Dyakonov-Shur theory~\cite{dyakonov_ieee_1996a} of photodetection of long-wavelength radiation assisted by plasmons whereby a resonantly-enhanced dc photovoltage appears between the source and drain contacts of a field-effect transistor hosting a two-dimensional electron gas in response to an oscillating electromagnetic field. In this theory, rectification occurs because of hydrodynamic nonlinearities. Here, we have shown that this intrinsic mechanism is much more general and occurs, with some intriguing qualitative differences, well beyond the frequency regime in which hydrodynamic theory applies.

The idea is that, on general grounds, the two-dimensional electron gas in the field-effect transistor channel has a nonlinear optical response of microscopic origin. We have shown that the latter yields a pseudo-Euler equation of motion for a collective velocity field, which is compactly reported in Eq.~(\ref{eq:general_E_j_rel}). Combining this new equation with the continuity equation one can calculate the dc photovoltage induced by an electromagnetic field well beyond the regime of frequencies where hydrodynamic theory applies. 

As a reality check, we have first used our formalism to derive the standard Euler equation of hydrodynamic theory---see Appendix~\ref{sect:hydro_cond_euler}---by employing recent results~\cite{sun_pnas_2018} on the second-order nonlinear conductivity of a hydrodynamic electron fluid.
We have then combined analytical and numerical work to illustrate differences between ordinary hydrodynamic theory and our pseudo-Euler formalism. For the sake of definiteness, we have used graphene as an example. For the latter material, our pseudo-Euler equation is reported in Eq.~(\ref{eq:euler_non_hydro_3}). Concrete results for the dc photovoltage one should expect in one- and two-dimensional photodetector geometries are presented and discussed in Sections~\ref{sect:photo_1D} and~\ref{sect:phot_2D}. Significant qualitative differences between our theory and conventional (i.e.~hydrodynamic) Dyakonov-Shur theory are clearly visible in Figs.~\ref{fig:three} and~\ref{fig:four} and, quite interestingly emerge both in the low- and high-frequency regimes, where, for the case of graphene, the natural frequency scale to keep in mind is the threshold for inter-band transitions, $\omega = 2E_{\rm F}/\hbar$, $E_{\rm F}$ being the Fermi energy.
These differences are amenable to experimental studies and greatly expand the knowledge of the fundamental physics behind intrinsic long-wavelength photodetection. 

\acknowledgements

This work has been sponsored by the European Union's Horizon 2020 research and innovation programme under grant agreement No.~785219---``Graphene Core2''. D.B. was supported by the Leverhulme Trust and the Russian Science Foundation - Grant 18-72-00234 (hydrodynamic theory). H.R. was supported by VR on Driven Quantum Matter, VILLUM FONDEN via the Center of Excellence for Dirac Materials (Grant No.~11744) and by KAW 2013.0096.

\appendix

\section{Derivation of the Euler equation from the hydrodynamic nonlinear conductivity}
\label{sect:hydro_cond_euler}
We start the reconstruction of the standard Euler equation by considering the hydrodynamic conductivity tensors~\cite{sun_pnas_2018}. The linear one reads
\begin{eqnarray} \label{eq:linear_cond}
\sigma_{ij}^{(1)}(\Omega) = \delta_{ij} \frac{i n_0 e^2}{\omega m}
~,
\end{eqnarray}
where $n_0$ is the equilibrium electron density and $m$ the electron mass. In the case of graphene, the mass $m$ in Eq.~(\ref{eq:linear_cond}) needs to be interpreted as an effective density-dependent mass---see main text and Ref.~\onlinecite{katsnelson_book}.
The second-order nonlinear conductivity in the hydrodynamic regime reads as following~\cite{sun_pnas_2018}
\begin{eqnarray} \label{eq:nonlinear_cond_hydro}
\sigma_{ijk}^{(2,{\rm Hy})}(\Omega, \Omega_1, \Omega_2) &=& \delta_{\Omega_1+\Omega_2,\Omega} \sum_{\beta} 
\big[ q_{1,\beta} d^{({\rm Hy})}_{ijk\beta}(\omega,\omega_1,\omega_2)
\nonumber\\
&+&
q_{2,\beta} d^{({\rm Hy})}_{ikj\beta}(\omega,\omega_2,\omega_1) \big]
~,
\end{eqnarray}
where $\delta_{\Omega_1+\Omega_2,\Omega} \equiv \delta_{{\bm q}_1+{\bm q}_2,{\bm q}} \delta_{\omega_1+\omega_2,\omega}$ and
\begin{eqnarray} \label{eq:d_hydro_def}
d^{({\rm Hy})}_{ijk\beta}(\omega,\omega_1,\omega_2) &=& d_0 \frac{\omega_1 \delta_{i\beta} \delta_{jk} + \omega \delta_{j\beta} \delta_{ik}}{\omega \omega_1^2 \omega_2}
~.
\end{eqnarray}
Here 
\begin{eqnarray}\label{eq:dzero_main}
d_0 &\equiv& 
\frac{n_0 e^3}{m^2}
~.
\end{eqnarray}

Plugging Eqs.~(\ref{eq:linear_cond})-(\ref{eq:nonlinear_cond_hydro}) into Eqs.~(\ref{eq:linear_resistivity_general})-(\ref{eq:nonlinear_resistivity_general}) we find
\begin{eqnarray}
\rho_{ij}^{(1)}(\Omega) = -\delta_{ij} \frac{i\omega m}{n_0 e^2}
\end{eqnarray}
and
\begin{eqnarray}
\rho_{ijk}^{(2,{\rm Hy})}(\Omega, \Omega_1, \Omega_2) &=& \delta_{\Omega_1+\Omega_2,\Omega}
\frac{im\omega}{n_0^2 e^3} \Bigg(\frac{q_i}{\omega} \delta_{jk}
+\frac{q_{1,j}}{\omega_1} \delta_{ik}
\nonumber\\
&+&
\frac{q_{2,k}}{\omega_2} \delta_{ij} \Bigg)
~.
\end{eqnarray}
Using these expressions, Eq.~(\ref{eq:general_E_j_rel}) becomes
\begin{eqnarray}  \label{eq:euler_hydro_1}
\frac{n_0 e^2}{m} {\bm E}(\Omega) &=& -i\omega {\bm j}(\Omega) - \frac{i \omega}{2 n_0 e} \sum_{\Omega_1} \Bigg\{\frac{{\bm q}}{\omega} \big[{\bm j}(\Omega_1)\cdot {\bm j}(\Omega_2)\big] 
\nonumber\\
&+& \frac{{\bm q}_1\cdot {\bm j}(\Omega_1)}{\omega_1} {\bm j}(\Omega_2) 
+
\frac{{\bm q}_2\cdot {\bm j}(\Omega_2)}{\omega_2} {\bm j}(\Omega_1) \Bigg\}
~.
\end{eqnarray}
Here it is understood that $\omega_2 = \omega - \omega_1$ and ${\bm q}_2 = {\bm q} - {\bm q}_1$. 

Using the continuity equation and the symmetry properties of Eq.~(\ref{eq:euler_hydro_1}) by exchange of $\omega_1 \leftrightarrow \omega_2$ and ${\bm q}_1 \leftrightarrow {\bm q}_2$ we rewrite it as
\begin{eqnarray} \label{eq:euler_hydro_3}
\frac{n_0 e^2}{m} {\bm E}(\Omega) 
&=&
-i\omega {\bm j}(\Omega) + \frac{1}{n_0 e} \sum_{\Omega_1} \Big\{-i {\bm q}_1 \big[{\bm j}(\Omega_1)\cdot {\bm j}(\Omega_2)\big]
\nonumber\\
&+&
i \omega_1 e \delta n(\Omega_1) {\bm j}(\Omega_2) +
i\omega_2 e \delta n(\Omega_1) {\bm j}(\Omega_2) \Big\}
~.
\nonumber\\
\end{eqnarray}
Fourier-transforming back to real space and time the previous equation we then get
\begin{eqnarray} \label{eq:euler_hydro_4}
\frac{n_0 e^2}{m} {\bm E}({\bm r},t) &=& \partial_t {\bm j}({\bm r}, t)
- \frac{1}{n_0 e} \Big\{ \sum_i j_i({\bm r}, t){\bm \nabla} j_i({\bm r}, t)
\nonumber\\
&+&
e \partial_t\delta n({\bm r},t) {\bm j}({\bm r}, t) 
+ e \delta n({\bm r},t) \partial_t {\bm j}({\bm r}, t) \Big\}
~.
\nonumber\\
\end{eqnarray}
We now further manipulate this expression. First, we rewrite the last term on the second line of Eq.~(\ref{eq:euler_hydro_4}) using that, to the order of nonlinearity to which we are working,
\begin{eqnarray}
\partial_t {\bm j}({\bm r}, t) = \frac{n_0 e^2}{m} {\bm E}({\bm r},t)
~.
\end{eqnarray}
Hence, after few straightforward manipulations, Eq.~(\ref{eq:euler_hydro_4}) becomes
\begin{eqnarray} \label{eq:euler_hydro_6}
\frac{e^2}{m} n({\bm r},t) {\bm E}({\bm r},t) &=& \partial_t {\bm j}({\bm r}, t)
- \frac{1}{n_0 e} \Big\{ \sum_i j_i({\bm r}, t){\bm \nabla} j_i({\bm r}, t) 
\nonumber\\
&+&
e \partial_t\delta n({\bm r},t) {\bm j}({\bm r}, t) 
\Big\}
~,
\end{eqnarray}
where $n({\bm r},t) \equiv n_0 +  \delta n({\bm r},t)$. Next, to rewrite this as the standard Euler equation, we introduce the velocity field ${\bm v}({\bm r},t)$ as
\begin{eqnarray}
{\bm j}({\bm r}, t) \equiv -e n({\bm r}, t) {\bm v}({\bm r}, t)
~.
\end{eqnarray}
Plugging this into Eq.~(\ref{eq:euler_hydro_6}), after few straightforward manipulations we get
\begin{equation} \label{eq:euler_hydro_7}
n({\bm r}, t) \partial_t {\bm v}({\bm r}, t) + n_0 v_i({\bm r}, t){\bm \nabla} v_i({\bm r}, t) = -\frac{e}{m} n({\bm r},t) {\bm E}({\bm r},t) 
~,
\end{equation}
where we retained terms up to second order in nonlinearities. Consistently with the order at which we are working, we can now: (i) replace $n_0 \to n({\bm r},t)$ in the last term on the left-hand side of Eq.~(\ref{eq:euler_hydro_7}), and divide everything by $n({\bm r}, t)$. Using well-known vector calculus identities, we rewrite Eq.~(\ref{eq:euler_hydro_7}) as
\begin{eqnarray} \label{eq:euler_hydro_8}
\partial_t {\bm v}({\bm r}, t) &+& \big[{\bm v}({\bm r}, t)\cdot {\bm \nabla}\big] {\bm v}({\bm r}, t) - \big[{\bm \nabla}\times {\bm v}({\bm r}, t)\big] \times {\bm v}({\bm r}, t) 
\nonumber\\
&=& - \frac{e}{m} {\bm E}({\bm r},t)
~.
\end{eqnarray}
This is almost Euler's equation, except for the third term on the right-hand side. We now manipulate it by using that, according to Faraday's equation,
\begin{eqnarray}
{\bm \nabla}\times {\bm E}({\bm r},t) = \frac{1}{c} \partial_t {\bm B}({\bm r},t)
~,
\end{eqnarray}
which allows to rewrite the time-derivative of the vorticity as
\begin{eqnarray}
\partial_t \big[{\bm \nabla}\times {\bm v}({\bm r}, t)\big] =
- \frac{e}{m c} \partial_t {\bm B}({\bm r},t)
~.
\end{eqnarray}
Integrating this equation and setting the time-independent constant equal to zero (which is consistent with the fact that no external time-independent magnetic field is present), and plugging the result in Eq.~(\ref{eq:euler_hydro_8}) we finally get
\begin{eqnarray} \label{eq:euler_hydro_final}
\partial_t {\bm v}({\bm r}, t) &+& \big[{\bm v}({\bm r}, t)\cdot {\bm \nabla}\big] {\bm v}({\bm r}, t) + \frac{e}{m c} {\bm B}({\bm r},t) \times {\bm v}({\bm r}, t) 
\nonumber\\
&=& - \frac{e}{m} {\bm E}({\bm r},t)
~,
\end{eqnarray}
which is the Euler equation in the presence of a Lorentz force due to the self-induced magnetic field ${\bm B}({\bm r},t)$.

\section{Simplification of the last term of Eq.~(\ref{eq:cont_NS_nonlinear_2}).}
\label{app:simplification_nonlinear}
In this Section we detail a few algebraic steps that we have carried out to manipulate the last term on the right-hand side of Eq.~(\ref{eq:cont_NS_nonlinear_2}). 

By using the second of Eqs.~(\ref{eq:cont_NS_poisson_linear}) we rewrite it as
\begin{widetext}
\begin{eqnarray} \label{eq:g_manip_1}
g(x,y) &\equiv& \Big\langle
\big[{\bm v}_1 ({\bm r},t) \times {\bm \nabla}\big] \times {\bm v}_1 ({\bm r},t)
+ 2\frac{e}{m} 
\big[{\bm \nabla}U_1 ({\bm r},t) \times {\bm \nabla}\big] \times {\bm \alpha}_1({\bm r},t)
\Big\rangle
\nonumber\\
&=&
\left\langle
\big[{\bm v}_1 ({\bm r},t) \times {\bm \nabla}\big] \times {\bm v}_1 ({\bm r},t)
+ 2
\left[ \left(\partial_t + \frac{1}{\tau}\right) {\bm v}_1({\bm r}, t) \times {\bm \nabla}\right] \times {\bm \alpha}_1({\bm r},t)
\right\rangle
~.
\end{eqnarray}
Using the fact that $\langle\ldots\rangle$ is an integral over time between $0$ and $2\pi/\omega$ (i.e.~an average over one cycle of the external oscillating field), we now integrate the second term on the last line of Eq.~(\ref{eq:g_manip_1}) by parts. Since all functions are periodic, boundary contributions cancel. We get
\begin{eqnarray} \label{eq:g_manip_2}
g(x,y) &=&
\left\langle
\big[{\bm v}_1 ({\bm r},t) \times {\bm \nabla}\big] \times {\bm v}_1 ({\bm r},t)
+ \frac{2}{\tau}
\left[ {\bm v}_1({\bm r}, t) \times {\bm \nabla}\right] \times {\bm \alpha}_1({\bm r},t)
- 2
\left[ {\bm v}_1({\bm r}, t) \times {\bm \nabla}\right] \times \partial_t {\bm \alpha}_1({\bm r},t)
\right\rangle
\nonumber\\
&=&
\left\langle
\frac{2}{\tau}
\left[ {\bm v}_1({\bm r}, t) \times {\bm \nabla}\right] \times {\bm \alpha}_1({\bm r},t)
-
\left[ {\bm v}_1({\bm r}, t) \times {\bm \nabla}\right] \times {\bm v}_1({\bm r},t)
\right\rangle
~.
\end{eqnarray}
Here we used that $\partial_t {\bm \alpha}_1({\bm r},t) = {\bm v}_1({\bm r},t)$. We now perform the integration. Using the definitions in Eq.~(\ref{eq:general_linear_solution_result}) we find
\begin{eqnarray} \label{eq:g_manip_3}
g(x,y) &=&
\frac{2}{\tau}
\big[ {\bm v}_1({\bm r}, \omega) \times {\bm \nabla}\big] \times {\bm \alpha}_1^\ast({\bm r},\omega)
+
\frac{2}{\tau}
\big[ {\bm v}_1^\ast({\bm r}, \omega) \times {\bm \nabla}\big] \times {\bm \alpha}_1({\bm r},\omega)
\nonumber\\
&-&
\big[ {\bm v}_1({\bm r}, \omega) \times {\bm \nabla}\big] \times {\bm v}_1^\ast({\bm r},\omega)
-
\big[ {\bm v}_1^\ast({\bm r}, \omega) \times {\bm \nabla}\big] \times {\bm v}_1({\bm r},\omega)
\nonumber\\
&=&
- \Bigg\{
\left( 1 + \frac{2 i}{\omega \tau}\right)  \big[ {\bm v}_1({\bm r}, \omega) \times {\bm \nabla}\big] \times {\bm v}_1^\ast({\bm r},\omega)
+
\left( 1 - \frac{2 i}{\omega \tau}\right) \big[ {\bm v}_1^\ast({\bm r}, \omega) \times {\bm \nabla}\big] \times {\bm v}_1({\bm r},\omega)
\Bigg\}
\nonumber\\
&=&
- \Big\{
\big[ {\bm v}_1({\bm r}, \omega) \times {\bm \nabla}\big] \times {\bm v}_1^\ast({\bm r},\omega) + \big[ {\bm v}_1^\ast({\bm r}, \omega) \times {\bm \nabla}\big] \times {\bm v}_1({\bm r},\omega)
\Big\}
\nonumber\\
&-&
\frac{2 i}{\omega \tau} \Big\{ \big[ {\bm v}_1({\bm r}, \omega) \times {\bm \nabla}\big] \times {\bm v}_1^\ast({\bm r},\omega) - \big[ {\bm v}_1^\ast({\bm r}, \omega) \times {\bm \nabla}\big] \times {\bm v}_1({\bm r},\omega)
\Big\}
~.
\end{eqnarray}
Here we used that
\begin{eqnarray}
{\bm \alpha}_1({\bm r},\omega) = i \frac{{\bm v}_1({\bm r},\omega)}{\omega}
~.
\end{eqnarray}
We now consider the two terms in Eq.~(\ref{eq:g_manip_3}) separately. We start from the second one, which can be rewritten as
\begin{eqnarray} \label{eq:g_manip_3_2}
g_2(x,y) &\equiv&
-\frac{2 i}{\omega \tau} \Big\{ \big[ {\bm v}_1({\bm r}, \omega) \times {\bm \nabla}\big] \times {\bm v}_1^\ast({\bm r},\omega) - \big[ {\bm v}_1^\ast({\bm r}, \omega) \times {\bm \nabla}\big] \times {\bm v}_1({\bm r},\omega)
\Big\}
\nonumber\\
&=&
-\frac{2 i}{\omega \tau} \Big\{ 
v_{1,i}({\bm r}, \omega) {\bm \nabla} v_{1,i}^\ast({\bm r},\omega) -  {\bm v}_1({\bm r}, \omega)\big[{\bm \nabla}\cdot{\bm v}_1^\ast({\bm r},\omega)\big]
- v_{1,i}^\ast({\bm r}, \omega) {\bm \nabla} v_{1,i}({\bm r},\omega) + {\bm v}_1^\ast({\bm r}, \omega)\big[{\bm \nabla}\cdot{\bm v}_1({\bm r},\omega)\big]
\Big\}
\nonumber\\
&=&
-\frac{2 i}{\omega \tau} \Big\{ 
\big[{\bm v}_{1}({\bm r}, \omega) \cdot {\bm \nabla}\big] {\bm v}_{1}^\ast({\bm r},\omega) -  {\bm v}_1({\bm r}, \omega)\big[{\bm \nabla}\cdot{\bm v}_1^\ast({\bm r},\omega)\big]
- \big[{\bm v}^\ast_{1}({\bm r}, \omega) \cdot {\bm \nabla}\big] {\bm v}_{1}({\bm r},\omega) + {\bm v}_1^\ast({\bm r}, \omega)\big[{\bm \nabla}\cdot{\bm v}_1({\bm r},\omega)\big]
\Big\}
\nonumber\\
&=&
-\frac{2 i}{\omega \tau} {\bm \nabla}\times \big[ {\bm v}_1^\ast({\bm r}, \omega) \times {\bm v}_1({\bm r},\omega) \big]
~.
\end{eqnarray}
In passing from the second to the third line we used that
\begin{eqnarray}
\big[{\bm \nabla} \times {\bm v}_1^\ast({\bm r},\omega)\big]\times  {\bm v}_1({\bm r},\omega) = \big[{\bm v}_1({\bm r},\omega)\cdot {\bm \nabla}\big] {\bm v}_1^\ast({\bm r},\omega) - v_{1,i}({\bm r}, \omega) {\bm \nabla} v_{1,i}^\ast({\bm r},\omega)  = 0
~,
\end{eqnarray}
and similarly for its complex conjugate.

Similarly,
\begin{eqnarray} \label{eq:g_manip_3_1}
g_1(x,y) &\equiv&
-\Big\{ \big[ {\bm v}_1({\bm r}, \omega) \times {\bm \nabla}\big] \times {\bm v}_1^\ast({\bm r},\omega) + \big[ {\bm v}_1^\ast({\bm r}, \omega) \times {\bm \nabla}\big] \times {\bm v}_1({\bm r},\omega) \Big\}
\nonumber\\
&=&
-\Big\{ 
v_{1,i}({\bm r}, \omega) {\bm \nabla} v_{1,i}^\ast({\bm r},\omega) -  {\bm v}_1({\bm r}, \omega)\big[{\bm \nabla}\cdot{\bm v}_1^\ast({\bm r},\omega)\big]
+ v_{1,i}^\ast({\bm r}, \omega) {\bm \nabla} v_{1,i}({\bm r},\omega) - {\bm v}_1^\ast({\bm r}, \omega)\big[{\bm \nabla}\cdot{\bm v}_1({\bm r},\omega)\big]
\Big\}
\nonumber\\
&=&
-\left\{ 
{\bm \nabla} \big[{\bm v}_{1}({\bm r}, \omega) \cdot {\bm v}_{1}^\ast({\bm r},\omega)\big] 
+ {\bm v}_1({\bm r}, \omega) \frac{i \omega U_1^\ast({\bm r},\omega)}{U_0}
- {\bm v}_1^\ast({\bm r}, \omega) \frac{i \omega U_1({\bm r},\omega)}{U_0}
\right\}
~.
\end{eqnarray}
In the last line we used that $-i \omega U_1({\bm r},\omega) + U_0 {\bm \nabla}\cdot {\bm v}_{1}({\bm r}, \omega) = 0$ and its complex conjugate. 
\begin{eqnarray} \label{eq:g_manip_3_1_1}
g_1(x,y)
&=&
-\Bigg\{ 
{\bm \nabla} \big[{\bm v}_{1}({\bm r}, \omega) \cdot {\bm v}_{1}^\ast({\bm r},\omega)\big] 
+ \frac{(i \omega - 1/\tau) U_1^\ast({\bm r},\omega)  {\bm v}_1({\bm r}, \omega) - (i \omega +1/\tau) U_1({\bm r},\omega) {\bm v}_1^\ast({\bm r}, \omega)}{U_0}
\nonumber\\
&+&
\frac{U_1^\ast({\bm r},\omega) {\bm v}_1({\bm r}, \omega) + U_1({\bm r},\omega) {\bm v}_1^\ast({\bm r}, \omega)}{\tau U_0}
\Bigg\}
\nonumber\\
&=&
-\Bigg\{ 
{\bm \nabla} \big[{\bm v}_{1}({\bm r}, \omega) \cdot {\bm v}_{1}^\ast({\bm r},\omega)\big] 
- \frac{e}{m U_0} \big[ U_1^\ast({\bm r},\omega) {\bm \nabla} U_1({\bm r},\omega) + U_1({\bm r},\omega) {\bm \nabla} U_1^\ast({\bm r},\omega) \big]
\nonumber\\
&+&
\frac{U_1^\ast({\bm r},\omega) {\bm v}_1({\bm r}, \omega) + U_1({\bm r},\omega) {\bm v}_1^\ast({\bm r}, \omega)}{\tau U_0}
\Bigg\}
\nonumber\\
&=&
-\Bigg\{ 
{\bm \nabla} \left[{\bm v}_{1}({\bm r}, \omega) \cdot {\bm v}_{1}^\ast({\bm r},\omega) + \frac{s^2}{U_0^2} U_1^\ast({\bm r},\omega) U_1({\bm r},\omega) \right] 
+ \frac{U_1^\ast({\bm r},\omega) {\bm v}_1({\bm r}, \omega) + U_1({\bm r},\omega) {\bm v}_1^\ast({\bm r}, \omega)}{\tau U_0}
\Bigg\}
~.
\end{eqnarray}
Therefore, Eq.~(\ref{eq:cont_NS_nonlinear_2}) reads
\begin{eqnarray} \label{eq:cont_NS_nonlinear_3_1}
&&
{\bm \nabla}\cdot \big[U_0 \langle {\bm v}_2 ({\bm r})\rangle + U_1^\ast({\bm r},\omega) {\bm v}_1({\bm r}, \omega) + U_1({\bm r},\omega) {\bm v}_1^\ast({\bm r}, \omega)   \big] = 0 
~,
\end{eqnarray}
and
\begin{eqnarray} \label{eq:cont_NS_nonlinear_3_2}
\frac{e}{m} {\bm \nabla} U_2({\bm r}) &=& 
\frac{1}{U_0 \tau} \Big\{U_0 {\bm v}_2({\bm r}) + \lambda \big[ U_1^\ast({\bm r},\omega) {\bm v}_1({\bm r}, \omega) + U_1({\bm r},\omega) {\bm v}_1^\ast({\bm r}, \omega) \big] \Big\}
\nonumber\\
&+&
{\bm \nabla} \left[(1+\lambda){\bm v}_{1}({\bm r}, \omega) \cdot {\bm v}_{1}^\ast({\bm r},\omega) + \lambda \frac{s^2}{U_0^2} U_1^\ast({\bm r},\omega) U_1({\bm r},\omega) \right] 
+ \frac{2 i \lambda}{\omega \tau} {\bm \nabla}\times \big[ {\bm v}_1^\ast({\bm r}, \omega) \times {\bm v}_1({\bm r},\omega) \big]
~.
\end{eqnarray}
Defining
\begin{eqnarray} \label{eq:phi_def}
\phi({\bm r}) \equiv \frac{e}{m} U_2({\bm r}) - (1+\lambda){\bm v}_{1}({\bm r}, \omega) \cdot {\bm v}_{1}^\ast({\bm r},\omega) - \lambda \frac{s^2}{U_0^2} U_1^\ast({\bm r},\omega) U_1({\bm r},\omega)
~,
\end{eqnarray}
Eq.~(\ref{eq:cont_NS_nonlinear_3_2}) reduces to
\begin{eqnarray} \label{eq:cont_NS_nonlinear_3_3}
{\bm \nabla} \phi({\bm r}) &=& 
\frac{1}{U_0 \tau} \Big\{U_0 {\bm v}_2({\bm r}) + \lambda \big[ U_1^\ast({\bm r},\omega) {\bm v}_1({\bm r}, \omega) + U_1({\bm r},\omega) {\bm v}_1^\ast({\bm r}, \omega) \big] \Big\}
+\frac{2 i \lambda}{\omega \tau} {\bm \nabla}\times \big[ {\bm v}_1^\ast({\bm r}, \omega) \times {\bm v}_1({\bm r},\omega) \big]
~.
\end{eqnarray}
\end{widetext}


\begin{thebibliography}{77}
\bibitem{otherparameters}
	The estimate has been obtained for a sample encapsulated in a $80-{\rm nm}$-slab of hBN placed on top of a perfect metallic gate.
\end{thebibliography}
\end{document}